\documentclass[journal]{IEEEtran}
\ifCLASSINFOpdf
\else
\fi
\usepackage{color}
\usepackage{enumerate}
\usepackage[cmex10]{amsmath} 
\usepackage{siunitx}
\usepackage{import} 

\usepackage{indentfirst} 
\usepackage{standalone}
\usepackage[section]{placeins} 
\usepackage{pdflscape}
\usepackage{tabularx}
\usepackage{algorithm} 
\usepackage{amssymb} 
\usepackage{microtype} 
\usepackage{soul} 
\usepackage{mathtools}
\usepackage{multirow}
\usepackage{verbatim}
\usepackage{graphicx}

\usepackage{pgfplots}

\usepackage{subfloat} 
\usepackage{verbatim}

\usepackage[utf8]{inputenc}
\usepackage[T1]{fontenc}
\usepackage{amsmath} 
\usepackage[shellescape,latex]{gmp} 
\usepackage[space]{grffile}
\usepackage{booktabs} 
\usepackage{xfrac}
\usepackage{tabularx}

\usepackage{cite}
\usepackage{adjustbox}

\usepackage{verbatim}
\usepackage{url}

\usetikzlibrary{shapes}

\usepackage{multicol, blindtext}    
\usepackage[]{color}
\usepackage{framed}
\usepackage{alltt}
\usepackage[T1]{fontenc}
\usepackage{array}
\usepackage{colortbl}
\usepackage{booktabs}
\usepackage{multirow}
\usepackage{eurosym}
\usepackage{listings}
\usepackage{ifthen}
\usepackage{pgfplots}
\usepackage{pgfplotstable}
\usepackage{pgfcalendar}
\usepackage{pgfgantt}
\usepackage{tikz}
\usetikzlibrary{shapes.geometric, arrows}

\usepackage{longtable}
\usepackage{amsfonts}
\usepackage{subfigure}
\usepackage{float}

\usepackage{mathrsfs}
\usepackage{algpseudocode} 
\usepackage{textcomp}
\usepackage{xspace}

\pgfplotsset{compat=1.16}

\usepackage{fmtcount} 

\begin{document}

\title{DRL-based Energy-Efficient Baseband Function Deployments for Service-Oriented Open RAN}

\author{Haiyuan Li, Amin Emami, Karcius Assis, Antonis Vafeas, Ruizhi Yang, \\ Reza Nejabati, Shuangyi Yan, and Dimitra Simeonidou

\thanks{Haiyuan Li, Amin Emami, Antonis Vafeas, Ruizhi Yang, Reza Nejabati, Shuangyi Yan and Dimitra Simeonidou are with the High Performance Networks Group, Smart Internet Lab, School of Computer Science, Electrical and Electronic Engineering, and Engineering Maths (SCEEM), Faculty of Engineering, University of Bristol$^1$, BS8 1QU, U.K. (e-mail: ocean.h.li@bristol.ac.uk).}
\thanks{Karcius Day R. Assis is with the Electrical and Computer Engineering Department, Federal University of Bahia (UFBA), Salvador-BA, Brazil and also as a visitor at University of Bristol, U.K.}
}

\maketitle
\vspace{-1cm}
\begin{abstract}
Open Radio Access Network (Open RAN) has gained tremendous attention from industry and academia with decentralized baseband functions across multiple processing units located at different places. However, the ever-expanding scope of RANs, along with fluctuations in resource utilization across different locations and timeframes, necessitates the implementation of robust function management policies to minimize network energy consumption. 
Most recently developed strategies neglected the activation time and the required energy for the server activation process, while this process could offset the potential energy savings gained from server hibernation. Furthermore, user plane functions, which can be deployed on edge computing servers to provide low-latency services, have not been sufficiently considered. In this paper, a multi-agent deep reinforcement learning (DRL) based function deployment algorithm, coupled with a heuristic method, has been developed to minimize energy consumption while fulfilling multiple requests and adhering to latency and resource constraints. In an 8-MEC network, the DRL-based solution approaches the performance of the benchmark while offering up to 51\% energy savings compared to existing approaches. In a larger network of 14-MEC, it maintains a 38\% energy-saving advantage and ensures real-time response capabilities. Furthermore, this paper prototypes an Open RAN testbed to verify the feasibility of the proposed solution.
\end{abstract}


\section{Introduction}
\label{intro}
\IEEEPARstart{T}{he}
dawn of fifth-generation (5G) mobile networks has ushered in a new era of connectivity, demanding a more agile radio access network (RAN) to deliver exceptional services for a myriad of use cases, such as ultra-reliable low latency communication (uRLLC), enhanced mobile broadband (eMBB), and massive machine type communications (mMTC) \cite{singh2020evolution, abdalla2022toward, tataria20216g}. 
Open RAN, an innovative industry-wide initiative, seeks to revolutionize the landscape in RAN by disaggregating RAN components into radio unit (RU), distributed unit (DU), and centralized unit (CU), with different split options\cite{d2022orchestran}. With 5G RAN functional splits, connections between RUs and user plane functions (UPFs) are divided into fronthaul, midhaul, and backhaul segments.
Open interfaces and standards defined in Open RAN and multi-access edge computing (MEC) facilities further facilitate the flexible deployment of these RAN functions close to users in achieving service customization and low latency applications. \cite{d2020sl, balasubramanian2021ric}. 
However, inevitable imbalances and variations in resource usage, driven by both spatial and temporal dimensions \cite{tranos2015mobile, li2022resilient}, and the growth of network scales and requests, may lead to service failure and significant energy waste in maintaining unnecessary servers. To harness the immense potential brought by Open RAN and MEC, an effective baseband function deployment strategy is required to integrate distributed DUs, CUs and UPFs into suitable MECs.

Numerous research has been dedicated to develop baseband function management strategies to handle server activation status and resource allocation with minimize power consumption. The baseband function management algorithms have been developed for different RAN architectures, including C-RAN \cite{tinini2017optimal, mijumbi2015server, gao2019deep, sigwele2020energy} and Open RAN \cite{gao2021deep, xiao2020service, xiao2021energy, zorello2022power, wang2022edge, klinkowski2020optimization}.
These algorithms can be further classified based on the employed techniques into optimization-based \cite{ gao2021deep, mijumbi2015server, tinini2017optimal}, heuristic-based \cite{harutyunyan2019latency, zorello2022power, xiao2021energy, klinkowski2020optimization} or DRL-based strategy \cite{gao2019deep, gao2021deep,xiao2021energy, wang2022edge} respectively.

In the context of centralized-RAN (C-RAN), Tinini \textit{et al.} pointed out that the baseband function deployment management is a  multi-dimensional bin-packing problem in essence and utilized MILP to study the most power-efficient BBU placement to accommodate network node demands \cite{tinini2017optimal}. Similarly, Mijumbi \textit{et al.} formulated a multi-objective optimization problem that considers practical constraints, such as latency and server capacity \cite{mijumbi2015server}. However, the intractable complexity of MILP prohibited practical deployments.
As an alternative, Gao \textit{et al.} proposed a DRL-based solution with offline executions that reduces reasoning delay, and minimizes bandwidth usage, transport latency, and the number of activated MECs hosting BBU functions \cite{gao2021deep}. Despite approaching the performance of MILP in a small-scale network, the action space of their solution expands exponentially with the growth of network size, leading to prohibitive complexity and training costs in large-scale networks. In contrast, to maintain scalability while reducing the number of required BBUs, Sigwele \textit{et al.} proposed simulated annealing and genetic algorithms under the constraints of service coverage and quality of service (QoS) \cite{sigwele2020energy}. Although this research has highly reduced the power cost in C-RANs,  heterogeneous requests and more complex components in Open RAN exacerbate the complexity of the multi-dimensional bin packing problem, necessitating other strategies to manage DU, CU and UPF deployment. 

Given the evolving complexities in Open RAN, Xiao \textit{et al.} designed a DRL-based baseband function placement and routing strategy, by modifying the DRL action from BBU selection to DU and CU placement \cite{xiao2020service}.  However, this approach does not address the issue of action space explosion, and each inference step can only handle a single request, thereby not addressing resource competition between requests.
In contrast, Zorello \textit{et al.} and Xiao \textit{et al.} proposed MILPs and heuristics,  which can determine the service function chains (SFCs) for all requests while minimizing network power consumption under latency and network resource constraints \cite{zorello2022power, xiao2021energy}. Nevertheless, those studies ignored the activation time of the hibernating servers, therefore, focused on peak power rather than the actual energy consumption when evaluating the algorithm performance. Additionally, few researchers have taken UPF into account alongside DU and CU as part of the baseband function chain \cite{xiao2020service, gao2021deep, joda2022deep}, whereas UPFs on MEC servers play a critical role in enabling users to access data networks for uRLLC services \cite{kekki2018mec, leyva2019framework}.

In summary, there are three crucial concerns that need to be addressed when designing effective policies for managing baseband functions. Firstly, previous algorithms for baseband function placement did not consider the inclusion of UPF alongside DU and CU. With edge computing, computational resources can be allocated closer to the user end, allowing for UPF placement on MEC to provide low-latency services. Secondly, current studies have failed to consider activation time which has resulted in inaccurate results and offsetting gains from hibernating servers. Lastly, it remains unclear how DRL-based solutions will handle the possible action space explosion in large networks. To overcome these obstacles, the main contributions of this paper are summarized below:
\begin{itemize}
   \item We develop a robust multi-agent DRL-based algorithm,  by incorporating resource limitations for DU, CU and UPF placement, multi-latency limitations from both fronthaul and end-to-end perspectives, as well as the activation time of servers on MECs.
  \item To address the action space explosion, we employ a multi-agent approach, assigning MECs on the network to multiple DRL agents for server activation strategy, further integrated with the application of heuristic to perform function placement and routing provisioning.
  \item Based on OpenDaylight \cite{medved2014opendaylight}, OpenStack \cite{sefraoui2012openstack} and Open Source Management and orchestration (OSM) \cite{openmano}, a programmable Open RAN testbed is developed to validate the importance of activation energy costs and the feasibility of our proposed algorithm in practical networks.
\end{itemize}

To the authors’ knowledge, this is the first time that multiple UPFs and the energy consumption of MEC activation have been taken into account in baseband function placement, with feasibility verification conducted on an Open RAN testbed.
The proposed algorithm is evaluated through simulations on 8-MEC and 14-MEC RAN networks.  
All on standby mode (ASM), which keeps servers awake, exhibits the highest energy consumption after 100s and 120s of idle time in both RAN networks. The results confirm the benefits of putting servers into hibernation and underscore the importance of designing effective baseband function deployment strategies. 
Moreover, in the 8-MEC RAN with 150s of network vacancy, the developed multi-agent DRL-based solution approaches the optimal MILP result, attaining remarkable energy savings exceeding 12\%, 33\%, 51\%, and 71\% when compared to power minimized deployment (PMD) \cite{xiao2021energy}, random allocation, greedy heuristic procedure (GHP) \cite{casazza2017securing}, and ASM, respectively. This performance remains consistent even in larger networks and maintains an 8\% and 38\% energy-saving advantage compared to PMD and GHP, respectively. In the end, the validity of the simulation outcomes and algorithm feasibility are verified through a series of measurements on MEC activation energy costs at three different hibernation levels, transmission delay between MECs, and MEC load power consumption.

The remainder of this paper is organized as follows. Section II introduces the concept of baseband function deployments in NG RAN. Section III and IV detail a MILP formulation and a multi-agent DRL-based algorithm. The simulation results are given in Section V. Section VI details the implementation of our testbed at Bristol and reports the feasibility evaluation. Finally, Section VII concludes the paper.

\section{Baseband Function Deployment Scenario in Open RAN}

In 4G networks, C-RAN architecture achieves significant advantages in improved resource utilization, energy efficiency, and network management by centralizing baseband processing into BBU. This centralized solution allows for better coordination among cells, dynamic allocation of resources, and reduced operational and capital expenditures due to consolidating equipment and infrastructure \cite{de2016overview, zhou2015energy}.
However, the advent of 5G networks, characterized by increased speed, reduced latency, and extensive device connectivity requirements, poses challenges for the C-RAN architecture. One such challenge is the potential latency issues that may arise due to the distance between the BBU and the remote radio unit. Moreover, while the centralized approach benefits resource sharing, it lacks the agility to address the diverse demands of 5G networks, in which use cases such as eMBB, mMTC, and uRLLC coexist. Additionally, the absence of open interfaces in traditional C-RAN architecture might restrict network deployment flexibility and hinders vendor interoperability \cite{gavrilovska2020cloud, niknam2022intelligent}.

In response, capitalizing on emerging edge computing and network function virtualization technologies, 3GPP and other standardization bodies proposed a novel concept, Open RAN, which includes DU, CU, and UPF that can be virtualized and tailored in distributed MEC servers to cater to heterogeneous requests \cite{ mollahasani2021dynamic}. Additionally, predefined open interfaces enable communication and collaboration between devices and systems from different manufacturers and operators, further enhancing service flexibility \cite{gavrilovska2020cloud}. The inherent modularity and open interface of Open RAN allow for rapid adaptation to dynamic 5G network demands and also facilitate seamless integration with a range of devices and technologies, providing operators with a streamlined transition from existing network frameworks \cite{bonati2020open}. Moreover, Open RAN empowers network operators with centralized and synchronized resource management capabilities, thereby reducing operation and management costs and improving service quality.

\begin{figure}[t]
    \centering
    \includegraphics[width=0.8\linewidth]{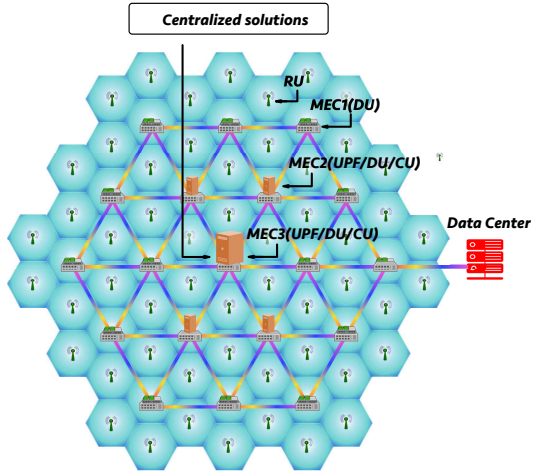}
    \caption{Baseband function deployment scenario in Open RAN}
    \label{scenario}
\end{figure}

A possible baseband function deployment scenario in Open RAN is shown in Figure \ref{scenario}. The DU demands significant resources to satisfy precise timing synchronization as specified in the Open RAN S-plane by IEEE 1588, as well as processing objectives such as the low-density parity check decoding defined in 3GPP 5G New Radio \cite{papatheofanous2021ldpc}. Consequently, many MEC servers may lack spare computing resources to host CU and UPF services influenced by deployment scenarios and can only cater to DU functions. Considering the computing resource differences, MECs are classified into three categories where MEC1 only provides DU service, MEC2 offers DU, CU, and UPF services, and MEC3 possesses the same functionalities as MEC2 but with more computing resources. 

Within Open RAN, requests from several RUs in a cell will be grouped and then served by SFCs comprising DU, CU, and UPF to access corresponding Date Network (DN) services. DUs, CUs, and UPFs from different cells can be placed on separate or identical MECs. The requests without latency constraints will directly go to the data center.
While considering three issues in previous studies, this baseband function placement process presents twofold challenges. The first challenge lies in allocating diverse baseband functions to different server types, ensuring service availability for multiple requests in dynamic networks while respecting computational, bandwidth, and latency constraints. The second challenge involves strategically activating the most suitable servers and locating baseband functions to minimize server energy consumption and maximize server resource utilization.
 
The following sections introduce a MILP-based solution and a multi-agent DRL-based algorithm to address these challenges by optimizing MEC activation, baseband function deployment, and routing provisioning for numerous requests.

\section{MILP Formulation for Baseband Function Deployments}

Based on the application scenario of baseband function deployment, a MILP-based solution is formulated to jointly optimize the MEC activation, DU/CU/UPF deployment and routing provisioning. The MILP-based solution provides an optimal solution and acts as a benchmark for the DRL-based solution. It is resolved using IBM ILOG CPLEX solver. 
Important notations used throughout this paper are summarized in Table~\ref{bigtable}.

\begin{table}[!t]
  \centering
  \caption{Notations Used Throughout the Paper.}
    \label{bigtable}%
    \scalebox{0.98}{
    \begin{tabular}{rlp{17.915em}}
    \toprule
          & \multicolumn{1}{c}{Notation} & \multicolumn{1}{c}{Description} \\
    \midrule
    \multicolumn{1}{l}{Index} 
        & $u$ & Activated MEC server \\
        & $s$ and $d$  & Source and destination of a set of requests \\
        & $i$ and $j$ & Origination and termination of a virtual link \\
        & $m$ and $n$  & Endpoints of a physical link \\
        & $f$ & Network function \\
    \midrule
    \multicolumn{1}{l}{Given} 
        & $c_{s}^{f}$  &  Computing requirement for $f$ from $s$   \\
        & $N$ & MEC set \\
        & $E$ & Link set  \\
        & $G = (N,E)$ & Network graph representation \\
        & $B_w$ & Bandwidth of physical link \\
        & $E_u$ & Activation energy of $u$ \\
        & $E_w$ & Unified switching energy cost \\
        & $G_{ u }^{f}$  &  Baseband function $f$ capacity of $u$ \\
        & $T^s$ & Data size of traffic from $s$  \\
        & $M_u$ & Traffic capacity of MEC $u$ \\
        & ${ d }_{ mn }$ & Distance between $m$ and $n$ \\
        & $l_{q}^{s}$ & Acceptable fronthaul distance of $s$\\
        & $l_{p}^{s}$  & Acceptable end-to-end distance of $s$ \\
        & $\chi$ & An arbitrarily large number \\
        & $\epsilon$ & An arbitrarily small number \\
        & ${\rho}_{ k }(f,g)$  & Function ordering indicator. It is 1 (resp. \text{-}1) if $f$ appears before (resp. after) $g$ \\
    \midrule
    \multicolumn{1}{l}{Variable} 
        & $\beta_u$ & Equals 1 if $u$ is activated  \\
        & ${{ v }^{ u }_{ij}}$  &  Virtual link arriving in $u$ \\
        & ${ \Lambda  }^{ sd }$ & Possible traffic flow from $s$ to $d$ \\
        & ${ { V }_{ ij } }$   &  Total aggregated traffic from $i$ to $j$  \\
        & ${ { l }_{ ij } }$   &  Total distance of an established virtual link from $i$ to $j$ \\
        & ${ \lambda  }_{ ij }^{ sd }$  & The component of $s$-$d$ node pair traffic offered from $i$ to $j$ \\
        & ${ B }_{ ij }^{ sd }$  & Equals 1 if ${ \lambda  }_{ ij }^{ sd }>0$; Equals to 0 if ${ \lambda  }_{ ij }^{ sd }=0$ \\
        & ${ P }_{ mn }^{ ij }$  &  The traffic that a virtual link from $i$ to $j$ passes through in link $m$-$n$ (integer) \\ 
        & ${ A }_{ mn }^{ ij }$  & Equals to 1 if ${ P }_{ mn }^{ ij }>0$; Equals 0 if ${ P }_{ mn }^{ ij }=0$  \\
        & ${ y }_{ fu}^{sd}$  & Equals 1 if function $f$ handles traffic $sd$ on node $u$ \\
        & ${ Y }_{ fu}^{ sd}$  & Equals 1 if $sd$ meets its assigned $f$ either before or on node $u$ \\
        & ${ { S }^{ sd } }$   &  Number of nodes switching when traffic goes from $s$ to $d$\\
        & ${ { b }^{ sd } }$   &  Equals 1 if there is traffic from $s$ to $d$ \\
    \bottomrule
    \end{tabular}%
}
\end{table}

\vspace{-0.5cm}

\subsection{MILP formulation}
This formulation aims to minimize the energy consumption in the network while satisfying all traffic requirements and resource limitations. Each optimization process handles multiple sets of requests from various MECs on the network, and each set of requests from an MEC is composed of individual requests received from RU with the same latency requirement. DU, CU and UPF are constrained by the computing resources in units of cores.
Consistent with \cite{xiao2021energy, zorello2022power}, the energy consumption of each server is assumed to increase linearly according to the size and amounts of the requests. The energy consumption over the network is therefore simplified to be determined by the activated MECs, function deployments, and routing paths, and the objective function for energy consumption after making these decisions can be written as: 
\begin{equation}
\label{eq_1}
\min \sum _{ u,s,d}(\beta_u E_{u}+S^{sd}E_{w})
\end{equation}
In this equation, $E_u$ and $\beta_u$ represent the activation energy cost of MEC $u$ and its activation decision. 
In addition, $S^{sd}$ is the switching times of requests $sd$ and $E_u$ is the unified switching energy cost. Overall, the objective function can be interpreted as achieving optimal energy cost through two joint policies of placement and routing to minimize the number of activating MEC1/2/3s and switching times.

The objective function is subject to constraints related to traffic demand, resource limitations, and latency requirements. Beginning with routing constraints for the virtual layer, Equation \ref{eq_2} and \ref{eq_novak1} represent the flow conservation constraints to handle traffic $T^{s}$ oriented from node $s$, where $\lambda_{i j}^{s d}$ stands for the traffic from the $sd$ pair on link $ij$ and $\Lambda^{s d}$ denotes possible traffic flow from s to d.
\begin{equation}
    \label{eq_2}
    \sum_j \lambda_{i j}^{s d}-\sum_j \lambda_{j i}^{s d}=\left\{\begin{array}{ll}
    \Lambda^{s d} & i=s \\ -\Lambda^{s d} & i=d \\ 0 & i \neq s, d
    \end{array}\right.
\end{equation}
\begin{equation}
    \label{eq_novak1}
    \sum _{ d }  \Lambda^{s d}=T^{ s } \quad \forall s
\end{equation}
Equations \ref{eq_4} and \ref{eq_5} impose total flow constraints on a virtual link in which $V_{ij}$ demonstrates the traffic on link $ij$ from all $sd$ pairs.
\begin{equation}
    \label{eq_4}
    \lambda_{ ij }^{ sd }\leq{ \Lambda  }^{ sd } \quad \forall s,d,i,j
\end{equation}
\begin{equation}
    \label{eq_5}
    \sum _{ sd } \lambda _{ ij }^{ sd }=V_{ ij } \quad \forall ij
\end{equation}
The indicator ${ B }_{ ij }^{ sd }$ for virtual links with traffic $sd$ can be acquired by
\begin{equation}
    \label{eq_6}
    { B }_{ ij }^{ sd } = \lceil\frac{ \lambda _{ ij }^{ sd }}{\chi}\rceil\quad \forall s,d,i,j
\end{equation}
where $\chi$ is an arbitrarily large number. 
${ B }_{ ij }^{ sd }$ equals 1 when ${ \lambda  }_{ ij }^{ sd }>0$, and ${ B }_{ ij }^{ sd }$ equals 0 when ${ \lambda  }_{ ij }^{ sd } = 0$.
Equations \ref{eq_7} and \ref{eq_8} illustrate the virtual links arriving at a destination node $d$ from any node $s$, accounting for their quantity. $v_{ij}^{u}$ denotes the virtual link arriving in $u$.
\begin{equation}
    \label{eq_7}
    \sum_{s}{ B }_{ij}^{sd} = v_{ij}^{u} \quad \forall s \neq u 
\end{equation}
\begin{equation}
    \label{eq_8}
    \sum_{ij}v_{ij}^{u} \leq M_u \quad \forall u;
\end{equation}

Moving on to the physical layer routing constraints, Equation \ref{eq_9}, similar to Equation \ref{eq_2}, expresses the flow conservation constraint of physical layer routing, wherein $P_{n m}^{i j}$ represents the traffic that a virtual link from $i$ to $j$ passes through in link $m$-$n$.
\begin{equation}
    \label{eq_9}
    \sum_n P_{m n}^{i j}-\sum_n P_{n m}^{i j}=\left\{\begin{array}{ll}
    V_{i j} & m=i \\
    -V_{i j} & m=j \\
    0 & m \neq i, j
    \end{array} \ \forall i, j, m\right.
\end{equation}
Equation \ref{eq_10} limits the bandwidth used not to exceed the capacity $B_w$ of physical links.
\begin{equation}   
    \sum _{ ij }{ P }_{ mn }^{ ij }\le B_{w} \quad \forall mn 
    \label{eq_10} 
\end{equation}
Equation \ref{eq_11} indicates the physical link $mn$ passing through the virtual link $ij$ and Equation \ref{eq_12} prevents traffic from being partitioned. ${ A }_{ mn }^{ ij }$ equals to 1 when ${ P }_{ mn }^{ ij }>0$, and it equals 0 when ${ P }_{ mn }^{ ij }=0$.
\begin{equation}
    \label{eq_11}
    { A }_{ mn }^{ ij }=\lceil\frac{ { P }_{ mn }^{ ij }}{\chi}\rceil \quad \forall i,j,m,n 
\end{equation}
\begin{equation}
    \label{eq_12}
    { A }_{ mn }^{ ij }+{ A }_{ ml }^{ ij }\le 1\quad \forall i,j,m \quad n\neq l
\end{equation}

Concerning latency requirements, the fronthaul and end-to-end latency constraints can be written as follows
\begin{equation}   
    \label{eq_13} 
    \sum _{ mn }^{  }{ A }_{ mn }^{ ij }d_{mn} = l_{ij}\quad \forall i,j
\end{equation} 
\begin{equation}
    \label{eq_14} 
    \sum _{ ij }^{  }{ B }_{ ij }^{ sd }l_{ij}\le l_{q}^s \quad   \forall j=u, \ DU \in u 
\end{equation}
\begin{equation}
    \label{eq_15} 
    \sum _{ ij }^{  }{ B }_{ ij }^{ sd }l_{ij}\le l_{p}^{s} \quad   \forall s,d=u, \ UPF \in u 
\end{equation}
where $l_{ij}$ denotes the total distance of an established virtual link from $i$ to $j$, and  $l_q^s$, $l_p^s$ represent the acceptable fronthaul and end-to-end distance of $s$, respectively. In addition, the carrying capability limitations of DU, CU and UPF are listed as
\begin{equation}
    \label{eq_25}
    \sum _{ s} c_s^f y_{ fu }^{ sd } \leq G_{ u }^{f} \quad \forall u, f=DU/CU
\end{equation}
\begin{equation}
    \label{eq_28}
\sum_{s}c_{s}^{f} { b }^{ sd } \leq G_{d}^{f}, \quad f = UPF
\end{equation}
${ b }^{ sd } $ equals 1 if there is traffic from $s$ to $d$, ${ y }_{ fu}^{sd}$ equals 1 if function $f$ handles traffic $sd$ on node $u$ and $G_{ u }^{f}$ denotes the baseband function $f$ capacity of $u$. Among them, ${ b }^{ sd } $ can be calculated by
\begin{equation}
    \label{eq_66}
    { b }_{ }^{ sd } = \lceil\frac{ \Lambda^{sd}}{\chi}\rceil\quad \forall s,d 
\end{equation}
Equation \ref{eq_27} counts the number of virtual links of the $sd$ pair and finds the number of switches that occurred from the $s$ to $d$ of this pair.
\begin{equation}
    \label{eq_27}
    \sum _{ ij } B_{ ij }^{ sd }-1 = S^{sd} \quad \forall s, d
\end{equation}

Regarding activation constraints, Equations \ref{eq_17}, \ref{eq_18}, and \ref{eq_19} affirm that all baseband functions find accommodation upon reaching the destination. ${ Y }_{ fu}^{ sd}$ equals 1 if $sd$ meets its assigned $f$ either before or on node $u$; otherwise, it equals 0.
\begin{equation}
    \label{eq_17}
    { Y }_{ fs }^{ sd} = 0\quad \forall s,d,f
\end{equation}
\begin{equation}
    \label{eq_18}
    { Y }_{ fd }^{ sd} = 1\quad \forall s,d,f
\end{equation}
\begin{equation}
    \label{eq_19}
    ({ B }_{ ij }^{ sd }-1)+{ Y }_{ fj }^{ sd}-{ Y }_{ f,i }^{ sd}\le { y }_{ fj }^{ sd }\quad \forall s,d,i,j,f
\end{equation}
However, a notable limitation of these equations is their inability to assign functions to the originating node, which contradicts our requirements. In response,  this formulation adopts auxiliary nodes connecting to every node as a virtual concept. It only helps the algorithm to calculate and place baseband functions on original nodes and will not be placed in real networks. A negligible latency from each auxiliary node to its adjacent original node and a huge energy cost is configured in the optimization process, allowing the auxiliary node to activate the original one and preventing itself from being activated.
Equation \ref{eq_20} accounts for the order of the baseband functions within the same chain. If a function $g$ appears before $f$ on the same SFC, then $\rho_k(f, g) = -\rho_k(g,f) = -1 $.
\begin{equation}
\label{eq_20}
{ Y }_{ fj }^{ sd}-{ Y }_{ gi }^{ sd}\geq { \rho }_{ k }( f, g)\quad \forall s,d,i,j,f;
\end{equation} 
Equation \ref{b1} $\beta_u$ is determined by $y_{fu}^{sd}$, which performs as indicator and equals 1 if function $f$ handles traffic $sd$ on node $u$.
\begin{equation}
    \label{b1}
    { y }_{ fu }^{ sd } = \beta_{u}\quad \forall s,d,u,f
\end{equation}

In summary, the constraints outlined in this section effectively address the traffic requirements, resource limitations, and latency requirements, encompassing a range of critical factors such as flow conservation, routing, latency, resource allocation, and activation.

\subsection{Complexity Analysis of the MILP}
Regarding the complexity of the problem, the DU/CU/UPF placement and routing are demonstrated to be a non-deterministic polynomial-time hardness (NP-hard) problem through a reduction from the bin-packing problem, as established by Xiao \textit{et al.} in \cite{xiao2021energy}. They proved this property for optimizing DU/CU deployment with bandwidth, latency and computing resource constraints.
In this paper, the problem is further complicated by the introduction of multiple UPFs, traffic decrease on the SFC and some other hard limitations.
Because of the transitivity of NP-Hardness, $P$ in our paper is also NP-hard.
In addition, Equations \ref{eq_14} and \ref{eq_15} contain products of variables, introducing non-linearity into the problem and making the network more complex. To linearize the problem, referring to \cite{bisschop2006aimms}, we replace the product of two variables with a single new variable, subject to specific constraints. For instance, considering ${ B }_{ ij }^{ sd }$ as a binary variable and $l_{ij}$ as a continuous variable bounded by 0 $\leq$ $l_{ij} \leq \chi$, a continuous variable $z$ is introduced to represent the product, i.e., $z= { B }_{ ij }^{ sd }l_{ij}$. However, merely introducing $z$ is insufficient and additional constraints \ref{eq_b4}, \ref{eq_b5}, \ref{eq_b6} and \ref{eq_b7} are also added to ensure that $z$ captures the value of ${B }_{ ij }^{ sd }l_{ij}$.
\begin{equation}
    \label{eq_b4}
    { z } \leq \chi{B }_{ ij }^{ sd }
\end{equation}
\begin{equation}
    \label{eq_b5}
    { z } \leq l_{ij}
\end{equation}
\begin{equation}
    \label{eq_b6}
    { z } \geq l_{ij}-\chi(1-{B }_{ ij }^{ sd })
\end{equation}
\begin{equation}
    \label{eq_b7}
    { z } \geq 0
\end{equation}

Given the constraints imposed by Equations \ref{eq_4}, \ref{eq_5}, \ref{eq_11}, \ref{eq_12}, among others, the time complexity of the MILP is $O(N^4)$. However, due to the NP-hard nature of the problem, its computational complexity may be significantly higher than this figure and the actual solving time may be influenced by the solving policies of the CPLEX solver.

\section{Multi-agent DRL-based Baseband Function Deployments}
\label{multiag}

While MILP can achieve optimal results for baseband function deployment in Open RAN, it has significant drawbacks. The inherent complexity of MILP leads to extensive computation times, making it unsuitable for real-time applications. Additionally, its deterministic nature lacks the flexibility to adapt to dynamic and uncertain network environments. Furthermore, flow conservation constraints in this formulation are not capable of accommodating applications that involve decreasing traffic along SFCs.
In response, a multi-agent DRL algorithm is developed to provide a baseband function deployment solution for real-time network management with dynamic traffic patterns. 
Unlike traditional clustering, classification, or regression models, the problem necessitates real-time feedback from the environment upon each action. This distinct requirement underscores the aptness of the agent-environment interaction paradigm inherent in DRL. Moreover, in comparison to other learning algorithms such as evolutionary and degrading methods, DRL capitalizes on experience replay, enabling agents to leverage past experiences for improved learning stability.

\begin{figure}[b]
    \centering
    \includegraphics[width=0.85\linewidth]{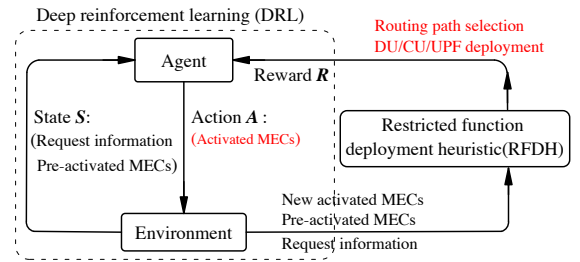}
    \caption{Overall DRL-based solution of baseband function deployment (DRL: MEC activation strategy; RFDH: DU, CU, UPF deployment and routing strategy).}
    \label{coop}
\end{figure}

In order to tackle the complexity of the SFC deployment problem and reduce the action space of intelligent agents, a novel two-step strategy is devised, which synergistically combines a DRL algorithm and a restricted function deployment heuristic (RFDH). As shown in Figure \ref{coop}, DRL agents collaborate, each making decisions on activating MECs while being aware of the choices of others, aiming to select the server combination with the lowest energy consumption for baseband placement. Distinctive from the common multi-agent DRL solutions, the collective decisions of all agents are then fed as inputs to a heuristic algorithm, RFDH, for further computation and bandwidth resource management. Leveraging the local optimization capabilities of heuristics and the adaptive learning capacity of DRL in dynamic environments, this strategy ensures superior optimization performance while maintaining adaptability to dynamic network environments. The following subsections introduce the multi-agent DRL-based algorithm and the RFDH algorithm.

\subsection{Multi-agent deep deterministic policy gradient based MEC activation strategy}

Building upon our innovative two-step strategy, we first model the network responses of function management to requests as a Markov Decision Process (MDP) \cite{bennett2013artificial}.
In the defined MDP, the discounted reward $R_t$ for a batch of tasks $Y$ can be given as:
\begin{equation}
R_{t}= \sum_{y=0}^{Y} \gamma^{y} r_{t+y}
\label{R_t}
\end{equation}
Here, $\gamma$ is the discount factor representing the importance of subsequent processing task rewards in $Y$ and indirectly expresses the degree of correlation between each step. $y$ is a natural number, and $r_{t + y}$ is the reward of a specific task, $t + y$. Based on Equation \ref{R_t}, the action value function is defined to evaluate action $a_t$ through the discounted return at state $s_t$:
\begin{equation}
Q_{\pi}\left(s_t, a_t\right)=\mathbb{E}\left[R_t \mid s_{t}, a_t\right]
\end{equation}
$\pi$ represents the policy function and corresponds to the probability density function of the action. Furthermore, another essential component, the state value function, is defined to illustrate the expected discounted return for selecting state $s_t$:
\begin{equation}
V_{\pi}\left(s_t\right)=\mathbb{E}_{a \sim \pi}\left[Q_{\pi}\left(s_t, a\right)\right]
\label{actcri}
\end{equation}
$V_{\pi}$ is employed to assess the quality of the policy function $\pi$. 

\begin{algorithm}[t]
\caption{{\small Multi-agent Deep Deterministic Policy Gradient (MADDPG) Algorithm}}
    \begin{algorithmic}[1]
        \small 
        \State Initialize replay memory $\mathcal{B}$ 
        \State Initialize the actor, target actor, critic and target critic with parameter $\theta_{1\sim L}, \theta_{1\sim L}^*, \omega_{1\sim L}, \omega_{1\sim L}^*$
        \For{$n$ in batch}
        \State Initialise the network resource state
        \State Get state $s_{n, 1\sim L}$
        \State Estimate $a_{n, 1\sim L}$ by $\pi(s_{n, 1\sim L} ; \theta_{n, 1\sim L})$
        \State Execute $a_{n, 1\sim L}$, get reward $r_{n, 1\sim L}$ by RFDH and next state 
        \Statex \quad \ \ $s_{n+1}$
        \State Store $(s_{n, 1\sim L}, a_{n, 1\sim L}, r_{n, 1\sim L}, s_{n+1, 1\sim L})$ in $\mathcal{B}$
        \State Get minibatch from $\mathcal{B}$
        \For{$m$ in minibatch}
        \State Estimate $a_{m+1, 1\sim L} $ by $ \pi(s_{m+1} ; \theta_{m, 1\sim L}^*)$
        \State Estimate $Q(s_{m, 1\sim L}, a_{m, 1\sim L} \mid \omega_{1\sim L})$
        \State Estimate $Q(s_{m+1, 1\sim L}, a_{m+1, 1\sim L} \mid \omega_{1\sim L}^*)$
        \State Use common reward for critics, $r_{t} = \overline{r_{t, 1\sim L}}$
        \State Perform gradient descent for critic based on TD algorithm
        \State Get $d_{\omega, \theta, t}=\frac{\partial Q(s_m, \pi(s_m ; {\theta}) ; \omega)}{\partial {\theta}}|_{\theta=\theta_{m, 1\sim L}, \omega=\omega_{m, 1\sim L}}$
        \State Perform gradient ascent for actor based on policy gradient 
        \Statex \qquad \ \ \ algorithm:
        \Statex \qquad \quad $\theta_{m+1, 1\sim L}=\theta_{m, 1\sim L}+\mu d_{\theta_{m, 1\sim L}, \omega_{m, 1\sim L}}$
        \State Update target network parameter based on 
        \Statex \qquad \quad $\theta_{1\sim L}^* = \tau * \theta_{1\sim L} + (1-\tau) * \theta_{1\sim L}^*$ 
        \Statex \qquad \quad $\omega_{1\sim L}^* = \tau * \omega_{1\sim L} + (1-\tau) * \omega_{1\sim L}^*$ 
        \EndFor
        \EndFor
    \end{algorithmic} 
    \label{maddpg}
\end{algorithm} 
\setlength{\textfloatsep}{0.7cm}

\begin{figure}[t]
    \centering
    \includegraphics[width=0.82\linewidth]{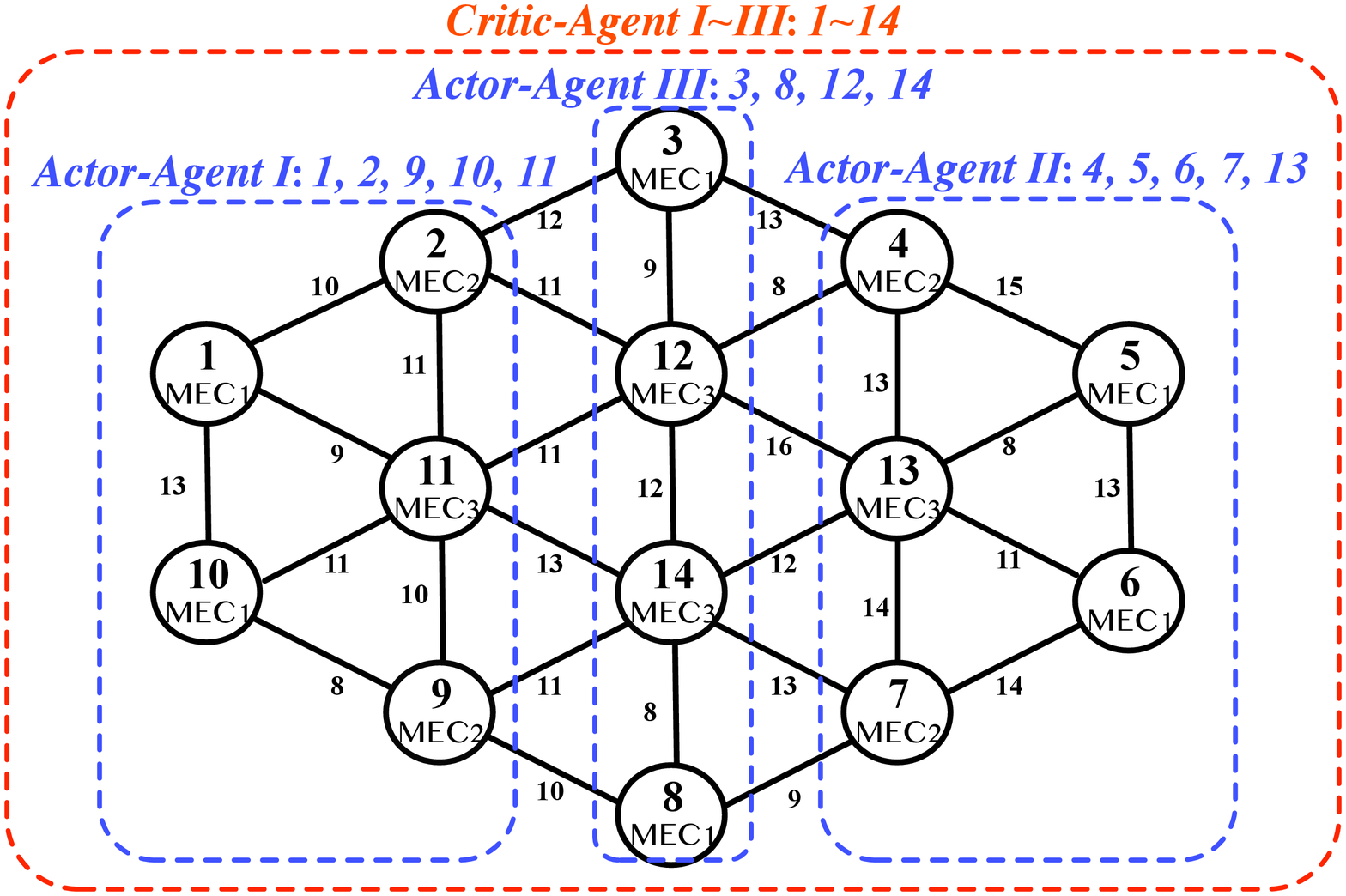}
    \caption{Example of MADDPG-based solution for baseband function deployment in a 14-MEC Network.}
    \label{network_example}
\end{figure}
This paper employs the Multi-agent Deep Deterministic Policy Gradient (MADDPG) to solve the proposed MDP. MADDPG is a promising DRL algorithm that utilizes policy gradient to estimate the maximum state value $V_{\pi}(s_t)$ \cite{silver2014deterministic}. Given that the state value function is equivalent to the expectation value of $Q_{\pi}$, as depicted in Equation \ref{actcri}, the fundamental principle of MADDPG involves employing two neural networks to approximate the policy function $\pi\left(a \mid s_t\right)$ and the action-value function $Q_{\pi}\left(s_t, a\right)$ to acquire $V_{\pi}(s_t)$. The interaction between policy and action-value functions is enabled by the actor-critic architecture \cite{lowe2017multi}.
Within this structure, the actor neural network employs the policy gradient algorithm to optimize the policy function, while the critic network utilizes the Temporal Difference (TD) algorithm \cite{mnih2013playing} to estimate the corresponding action-value function. 

To diminish the action space in the proposed resolution, all servers are split into $L$ groups, with each group corresponding to a DRL agent. The \textit{State} components of each actor and critic consist of the pre-activation status of $N$ servers and the parameters of each request, including fronthaul delay $l_q^s$, end-to-end delay $l_p^s$, initial data size $T_s$, and UPF resource requirement $c_s$. 
Based on the state information, the continuous \textit{Action} of each agent is discretized, which allows for determining the activation status of MECs within their designated groups.
It is worth noting that the fully cooperative (FC) scheme is adopted for model training, wherein $L$ critics share a common reward to optimize the cooperation between agents and minimize the activation energy consumption across the entire network. 
The reward and RFDH will be discussed in the following subsection.
Overall, the MADDPG algorithm is demonstrated in the Algorithm \ref{maddpg}. $\tau$ in the policy gradient algorithm is the coefficient to update the target network parameter. $\mu$ denotes the learning rate of DRL. $\theta$, $\theta^*$, $\omega$ and $\omega^*$ are parameters of actor, target actor, critic, and target critic, respectively.

The complexity of the DRL-based algorithm is analyzed as follows. Considering a $N$ node network, the action space of each actor $l$ is $2^{M_l}$, and the state space size is $5 * N$. $M_l$ is the number of MECs in the group $L$ and $\textstyle \sum_l^L M_l = N$. Based on the actor-critic architecture, we can acquire that the state space size of each critic is $L * 2^{M_l} + 5 *  N$. In each critic, the output with shape $1*1$ is used as the criterion of the policy gradient algorithm for the corresponding actor. 
Taking a 14-node network shown in Figure \ref{network_example} as an example, we can assign nodes 1, 2, 9, 10, 11 to actor agent I, nodes 4, 5, 6, 7, 13 to actor agent II and the rest to actor agent III. The action and state space of the three actors are 32, 32, 16 and 70, 70, 70, respectively. Since the entire network information and actions done by three actors are input to three critics as state information, the state space of each critic is 150. 
With the increase in network size $N$, to realize the scalability, the DRL-based algorithm can maintain an acceptable action space size $2^{M_l}$ with a larger number of agent $L$.

\subsection{Restricted function deployment heuristic (RFDH)}
\label{H}  
Restricted by the DRL actions (MEC activation status) in each DRL step, as shown in Algorithm \ref{RFDH}, we propose a heuristic RFDH to firstly determine the baseband function deployment and routing paths and secondly calculate the corresponding reward and feed it back to the DRL agent.

RFDH compiles all activated MECs into a set $\mathcal{K}$. It is important to note that MECs not belonging to the set $\mathcal{K}$ are prohibited from being activated. Within the heuristic, UPF, DU, and CU are deployed on the $\mathcal{K}$ in sequence, adhering to latency and network capacity constraints. 
When distributing UPF and DU, RFDH sorts the accessible MECs in descending order based on the number of requests they can serve. This approach aims to satisfy the most requests with the least resources. In addition, it sorts the deployment decision priority of the requests in ascending order according to their selection space size to accommodate more requests within the network. For instance, in a network where nodes A and B are pre-activated and requests from nodes C, D, and E seek UPF service. Assuming all these requests can access node A, while only requests from nodes D and E can access node B, traffic originating from node C will be accorded a greater priority for servicing by node A, given that it possesses a singular option. This ensures that limited resources will not be wasted by inappropriate resource allocations.

Regarding the path provisioning process, the Depth-First Search (DFS) algorithm \cite{tarjan1972depth} is adopted to traverse all available paths, subject to the end-to-end delay $l_p^s$. Within the optional routing space and subject to bandwidth limitation, routing path provisioning prioritizes fewer hops to reduce switching energy consumption. On the planned path, CU is prioritized to be placed on servers that are already activated and have sufficient resources.

\begin{algorithm}[t]
\caption{{\small Restricted Function Deployment Heuristic (RFDH)}}
    \begin{algorithmic}[1]
        \small
        \State Store new activated and pre-activated MECs into list $\mathcal{K}$
        \State Deploy UPFs on MECs in $\mathcal{K}$ based on latency and network capacity constraints
        \State Calculate and store traffic demands and accessible MECs in dictionary $\mathcal{S}_0$
        \State Sort MEC2/3 in $\mathcal{K}$ in descending order based on the number of requests they can serve
        \State Sort requests in Values in $\mathcal{S}_0$ in ascending order based on the number of MECs they can access
        \For{MEC2/3 $u$ in $\mathcal{K}$, Sorted traffic $s$ in $\mathcal{S}_0[u]$}
        \State Deploy UPF on $u$ for $s$ if resources are sufficient, otherwise 
        \Statex \quad \ set DRL reward = -1
        \EndFor
        \State Find feasible paths from origins to destinations based on $l_q^s$, store into dictionary $\mathcal{D}_0$
        \State Deploy DUs and CUs on activated MECs in $\mathcal{K}$ for demands $s$
        \State Calculate and store traffic demands and acceptable MECs and paths in dictionary $\mathcal{S}_1$
        \For{rest traffic $s$ in Key list of $\mathcal{S}_1$ without DU}
        \State Sort paths in $\mathcal{S}_1[s]$ based on the number of hops
        \For{all activated MEC1/2/3 $u$ in $\mathcal{K}$}
        \For{paths in $\mathcal{S}_1[s]$}
        \State Deploy DU on $u$ for $s$ if resources and bandwidth 
        \Statex \qquad \qquad \ \  constraints are met
        \State Store the used paths into dictionary $\mathcal{D}_1$
        \EndFor
        \EndFor
        \EndFor
        \State Deploy CUs on activated MECs in selected paths from $\mathcal{D}_1$
        \If{left no requests without CU}
        \State DRL reward $ r_t$ = $w_2/w_1$, otherwise $ r_t = -1 $
        \EndIf
    \end{algorithmic}
    \label{RFDH}
\end{algorithm}

Based on baseband function deployment decisions from DRL and RFDH, RFDH then calculates the corresponding reward for training purposes.
The \textit{Reward} is assigned a value of -1, acting as a penalty, should any remaining requests have not been serviced.
Otherwise, it equals $w_2/w_1$, where $w_1$ is the energy consumption determined by the DRL-based algorithm for routing and activating MECs, and $w_2$ is the energy cost when all the servers are activated. This reward design is consistent with the objective function in Equation \ref{eq_1}, providing higher rewards for actions that use less energy, thus minimizing energy consumption while satisfying all requests over the network.

The complexity of the RFDH is calculated as follows. 
Under the assumption of $m$ activated MEC1s and $J$ activated MEC2s and MEC3s, the complexity of distributing UPF services from activated MECs in $\mathcal{K}$ is $O(NJ + JlogJ + NlogN - J^2)$, where Merge Sort \cite{cole1988parallel} is used as the sorting algorithm. Based on the distributed MECs, we applied the DFT to find all the qualified paths with the time complexity of $O(N^2+NJ)$. After that, the complexity of utilizing $(J + m)$ activated MECs for DU and CU service can be represented by $O( N + (N-J-m)S_1logS_1 + (N-J-m)(J+m)S_1) $, where $S_1$ is the average length of Values in dictionary $\mathcal{S}_1$ in Algorithm \ref{RFDH}. Overall, the time complexity of RFDH can be written as $O(NJ - J^2 + JlogJ + NlogN + N + (N-J-m)S_1logS_1 + (N-J-m)(J+m)S_1)$. 

In summary, multi-agent DRL and RFDH integration offers a compelling approach for managing network resources and services, especially in large-scale and dynamic Open RAN environments. 
This technique renders the algorithm remarkably robust and resilient to changes in the network environment, showcasing its versatility and applicability in contemporary 5G systems.

\section{Simulation Results}
\label{results}

In this section, the setup information of Open RAN components and DRL-based algorithms is provided. Restricted by the scalability of MILP, the energy-saving performance of the several function deployment algorithms is first compared in a small network consisting of 8 MECs. Subsequently, their performance is explored in a larger network with 14 nodes and 29 links. The time and space complexity are calculated to demonstrate the training cost of DRL-based algorithms.

\subsection{Simulation Setup}
\begin{figure}[!t] 
        \setlength{\belowcaptionskip}{-0.2cm}
	\centering  
	\subfigure[Small network]{
	    \label{small}
		\includegraphics[width=0.4\linewidth]{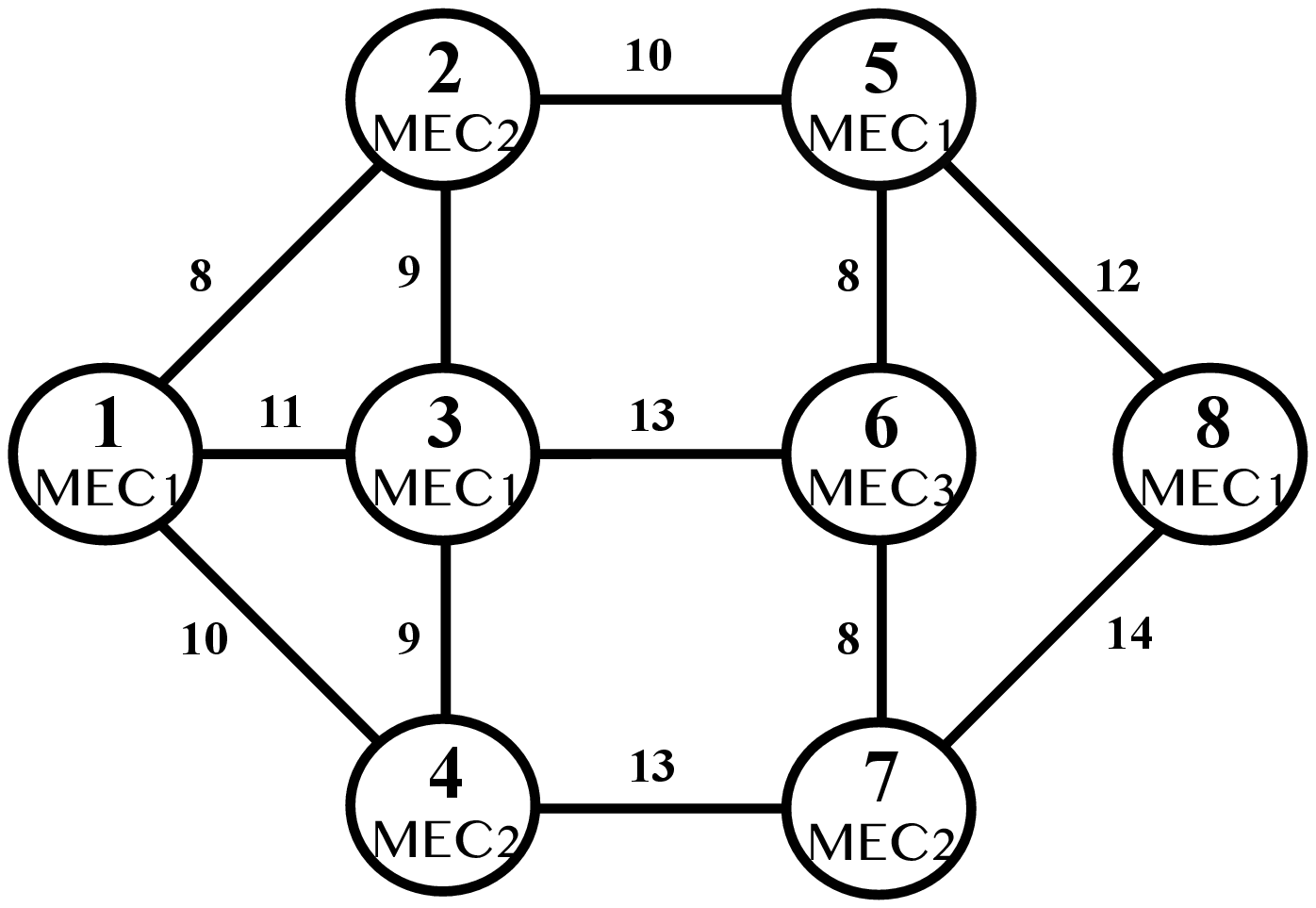}}
	\subfigure[Large network]{
	    \label{large}
		\includegraphics[width=0.54\linewidth]{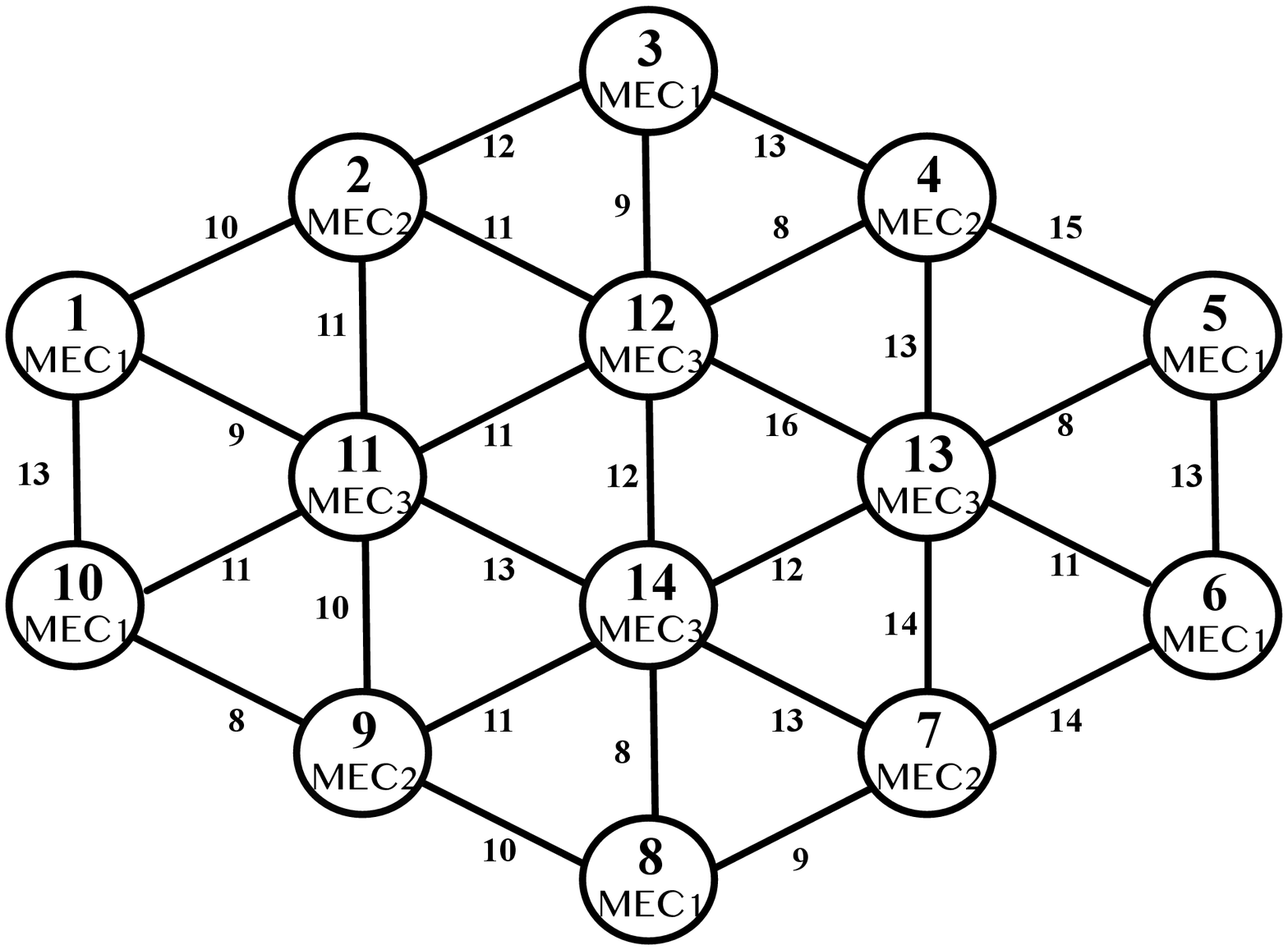}}
	\caption{Simulation networks.}
	\label{two-network}
\end{figure}

The simulated MEC networks with 8 and 14 servers are shown in Figures \ref{small} and \ref{large}. All MECs are provisioned with DU and CU capacities set at 50 cores. Specifically, MEC2 is equipped with UPF resources of 32 cores, while MEC3 boasts 50 cores for UPF.
Each link has a bandwidth of 50 Gbit/s. The operation power consumption for MEC1, MEC2, and MEC3 are 100W, 170W, and 200W, respectively. The corresponding activation power and time costs are 500W and 20s for MEC1 (i.e. $E_u = 10kJ$), 600W and 25s for MEC2 (i.e. $E_u = 15kJ$) and 700W and 30s for MEC3 (i.e. $E_u = 21kJ$). The switching energy consumption is uniformly set to 0.5 kJ. These parameters refer to the resource information and the server activation energy consumption detailed in the next section. For each set of requests from a MEC server, the data size, DU, and UPF resource requirements are randomly distributed between 8 Gbit/s to 12 Gbit/s, 8 cores to 12 cores, and 3 cores to 5 cores, respectively. 
Following the xhaul latency requirement proposed by IEEE standards association \cite{xiao2021energy, alam2018xhaul}, the fronthaul delay ranges from 11 km to 21 km, while the end-to-end delay is set between 28 km and 43 km. The computing requirement decreasing ratio along the function chain is configured as 0.2.

\subsection{8-MEC network}

\begin{figure}[b] 
	\centering  
        \subfigure[MILP for $F$ in 8-MEC network]{
	    \label{milp-8}
		\includegraphics[width=0.45\linewidth]{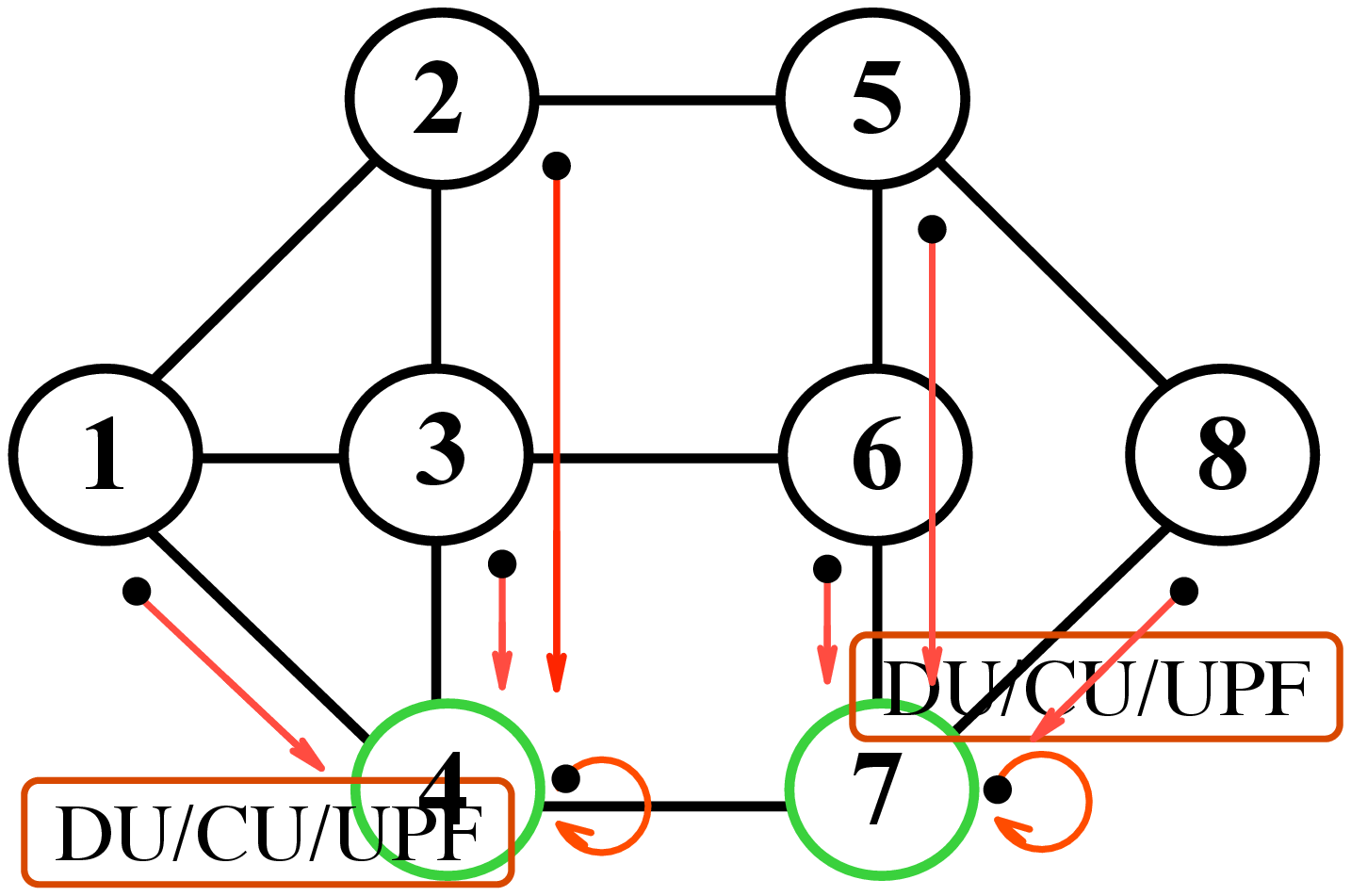}}\\
	\subfigure[DRL for $F$ in 8-MEC network]{
	    \label{drl-8}
		\includegraphics[width=0.45\linewidth]{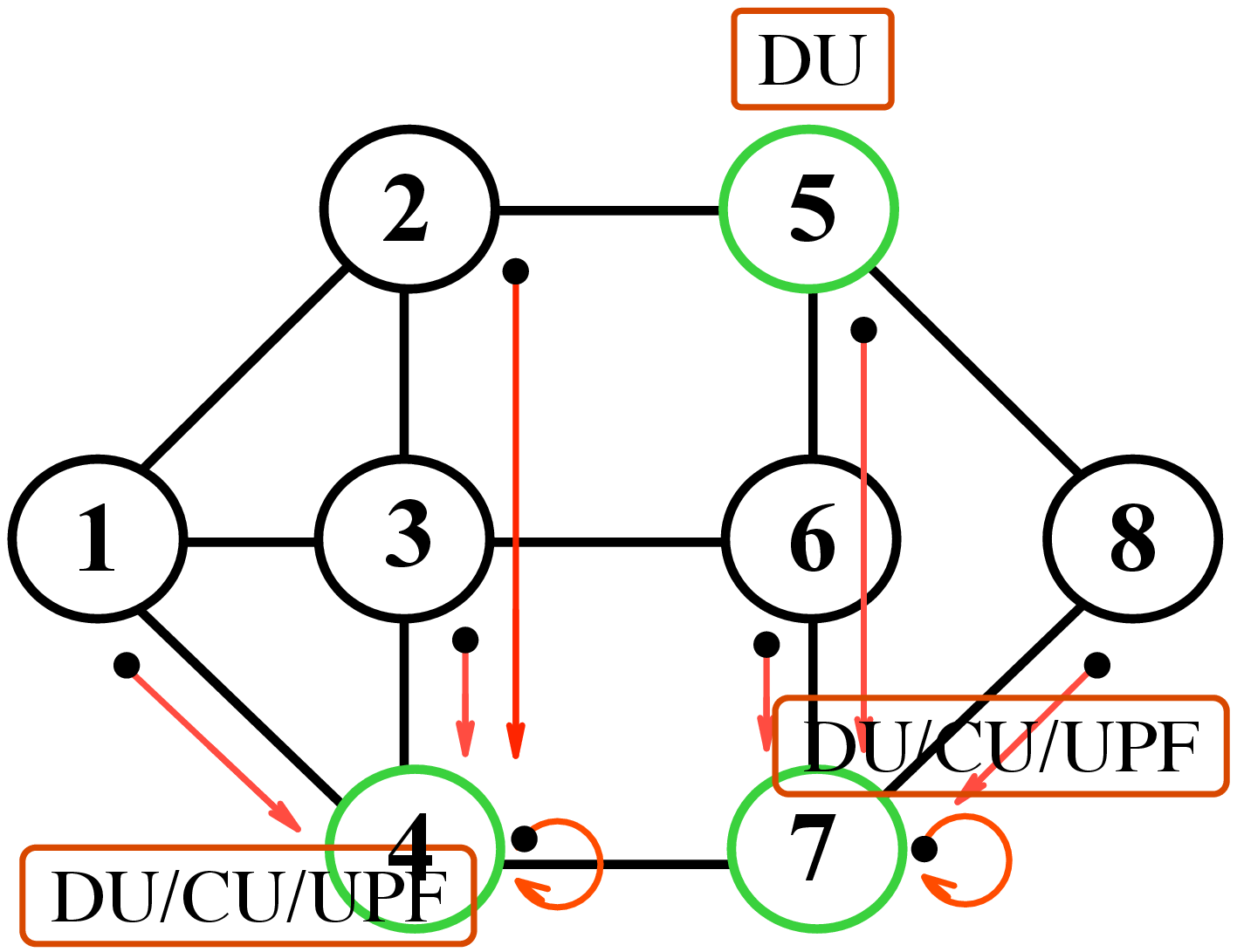}}
        \subfigure[DRL for $T$ in 14-MEC network]{
	    \label{drl-14}
		\includegraphics[width=0.45\linewidth]{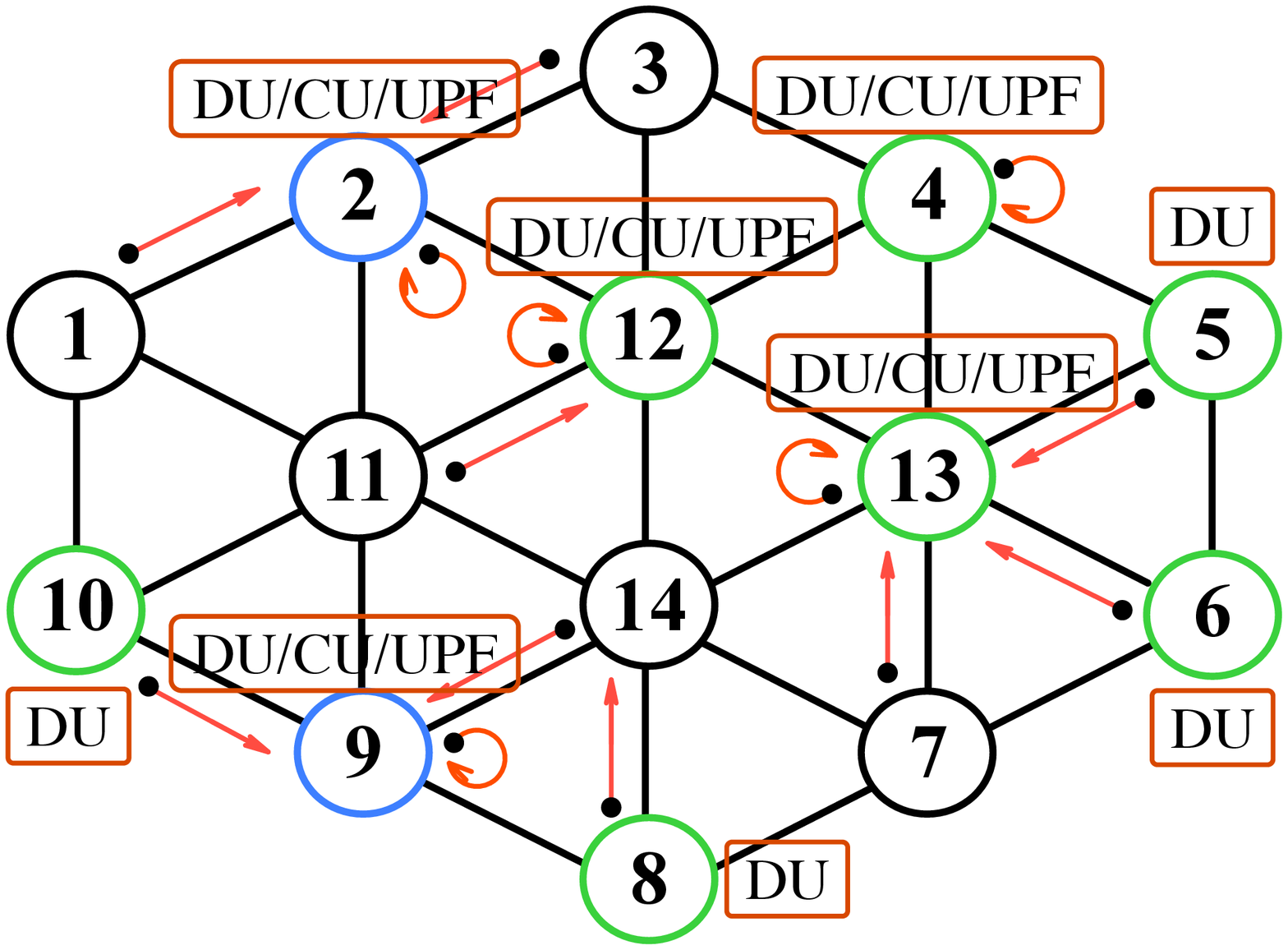}}
	\caption{Baseband function placement and path provisioning of MILP and MADDPG for traffic $F$ and $T$ in two networks.}
	\label{placement}
\end{figure}

To demonstrate the effectiveness of our designed algorithms, we first present the location of activated MECs, provisioned functions, and routing paths for traffic set $F$ in Figure \ref{milp-8} and \ref{drl-8}. The 8 requests in $F$ consist of fronthaul delay [14, 20, 17, 19, 19, 17, 17, 17] in km, end-to-end delay [28, 38, 40, 29, 41, 32, 34, 42] in km, data size [11, 9, 10, 9, 9, 8, 10, 11] in Gbit/s, and computing resource requirements for DU and UPF [11, 9, 10, 9, 9, 8, 10, 11] and [5, 5, 5, 5, 4, 3, 5, 4] in cores. Activated MECs and routing paths are marked by green circles and red arrows.

\begin{figure*}[t] 
\setlength{\belowcaptionskip}{-0.2cm}
 \centering  
	\subfigure[Traffic $F$ in the 8-MEC network]{
	    \label{vacant-8}
		\includegraphics[width=0.236\linewidth]{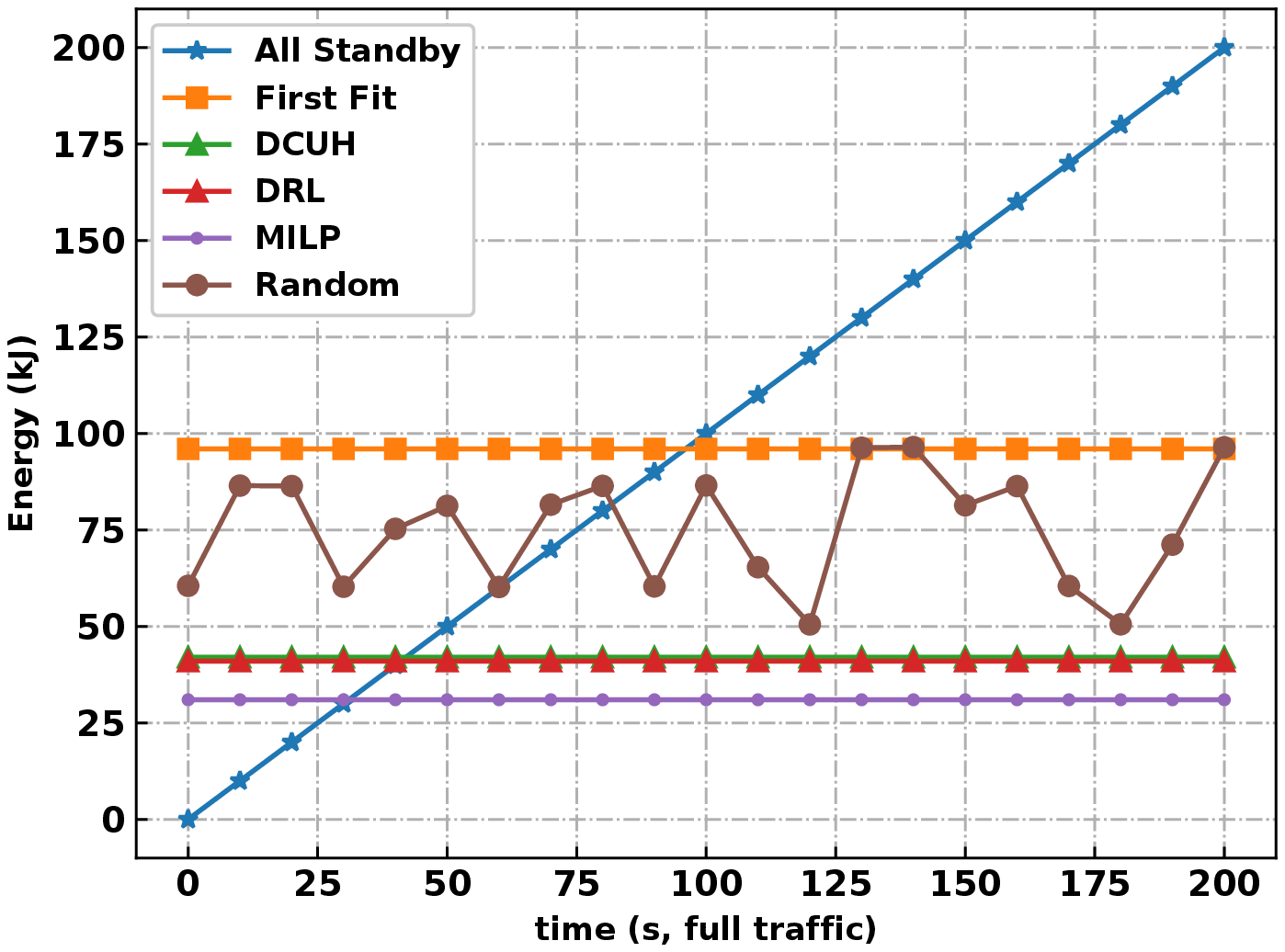}}
	\subfigure[50 random requests with 150s of network vacancy in the 8-MEC network.]{
	    \label{150-8}
		\includegraphics[width=0.236\linewidth]{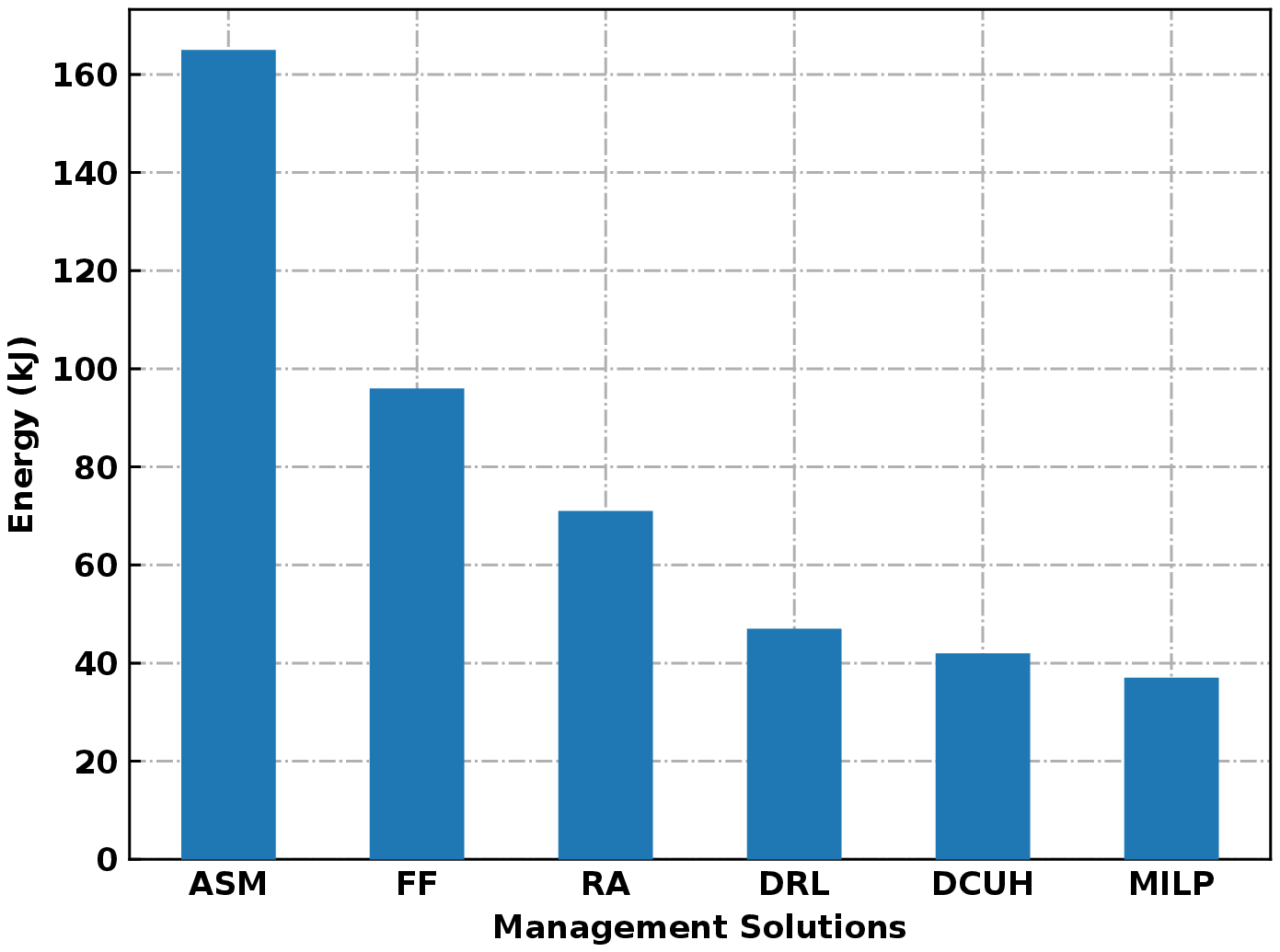}}
        \subfigure[Traffic $T$ in the 14-MEC network]{
	    \label{vacant-14}
		\includegraphics[width=0.236\linewidth]{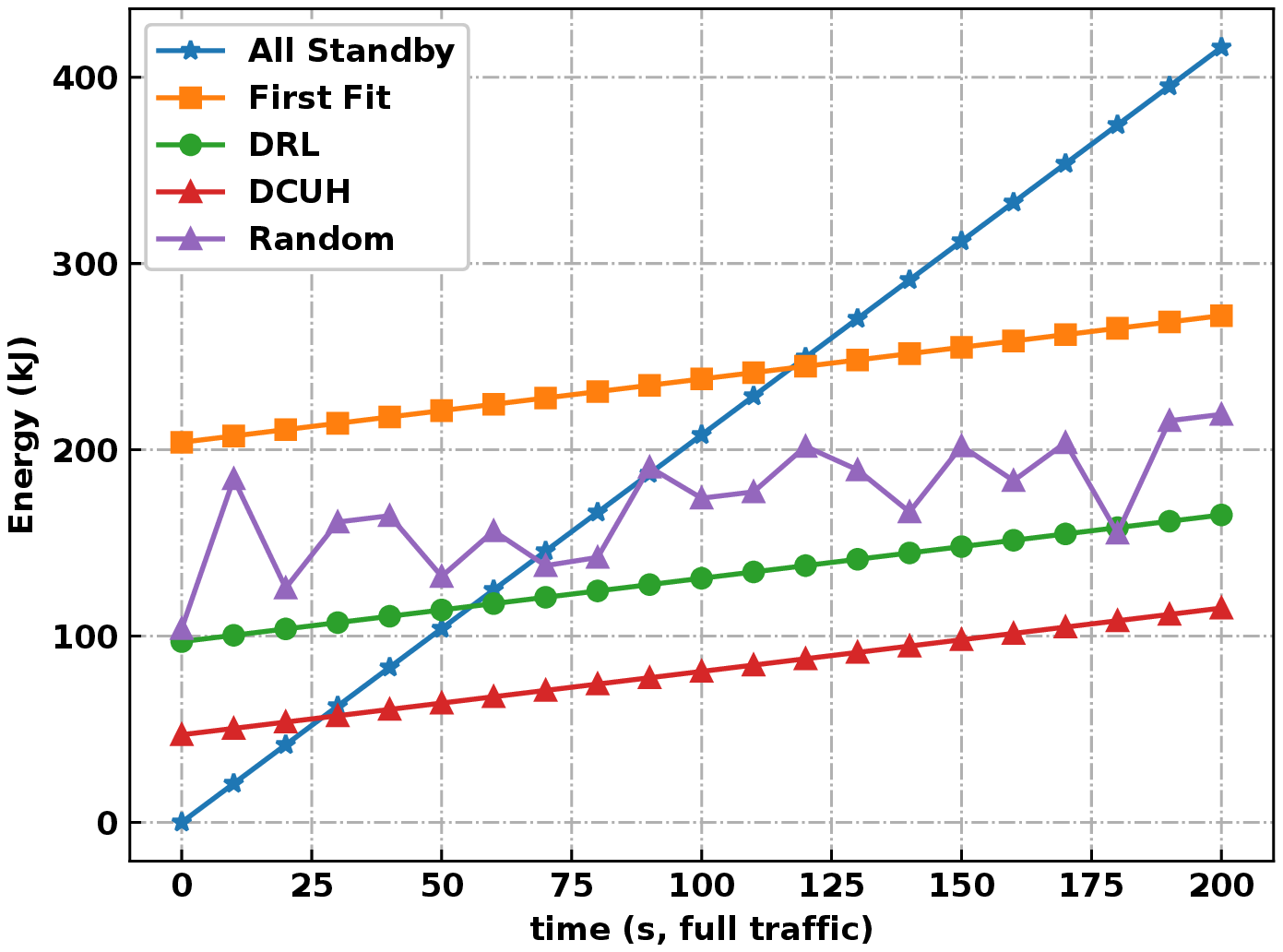}}
	\subfigure[50 random requests with 150s of network vacancy in the 14-MEC network.]{
	    \label{150-14}
		\includegraphics[width=0.236\linewidth]{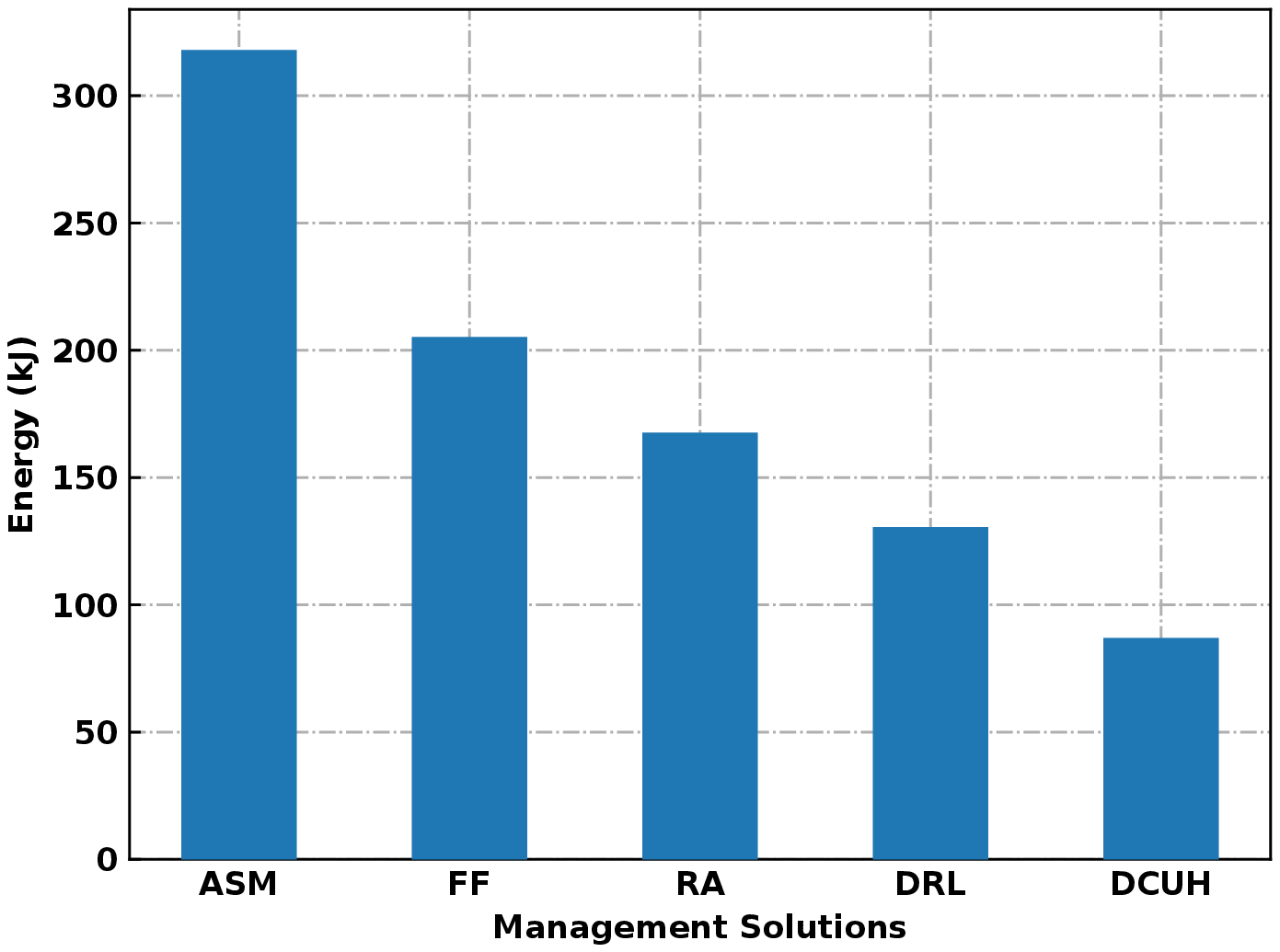}}
	\caption{Energy consumption comparison between different joint baseband function deployment algorithms in two networks.}
	\label{energy-8}
\end{figure*}

Moreover, the energy usage for deploying baseband functions for traffic $F$ with different management strategies is analyzed as the network idle time increases. This can be observed in Figure \ref{vacant-8}.
Blue, orange, purple, red, and green lines represent the requests for ASM, GHP \cite{casazza2017securing}, random allocation (RA), PMD \cite{xiao2021energy}, MILP and DRL, respectively. Among these, ASM keeps all MECs awake without activation cost, regardless of the request status. GHP is a priority-based policy proposed in \cite{casazza2017securing}, which selects servers with sufficient remaining resources in a greedy manner, RA randomly selects resources, ensuring all requests are fulfilled and PMD is a power-efficient heuristic that emphasizes deploying baseband functions on pre-activated servers to reduce activation cost. \cite{xiao2021energy}.
Initially, ASM was the most energy-efficient solution. However, it becomes the least efficient when the idle time exceeds 100s. 
This observation validates the energy-saving benefits of hibernating MECs during extended idle periods. 
The results of MILP are considered a benchmark due to its characteristics. Among the remaining strategies, the DRL-based strategy exhibits outstanding energy-saving performance. 
To obtain a generalized performance comparison, in Figure \ref{150-8}, we summarize the energy consumption of these solutions for 50 random requests with different latency and computing requirement settings, observed during a 150s network idle period. The DRL-based solution can save more than 12\%, 33\%, 51\%, and 71\% energy compared to PMD, RA, GHP, and ASM, respectively.
It is worth noting that although RA achieves better performance than GHP, its results are obtained under the assumption of satisfying all requests, which cannot be guaranteed in practical applications.

\begin{figure}[t]
    \centering
    \includegraphics[width=0.79\linewidth]{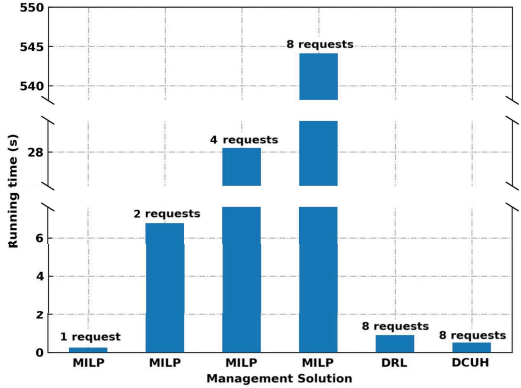}
    \caption{Running time comparison in the 8-MEC network.}
    \label{time-cost}
\end{figure}

\begin{figure}[t]
    \centering
    \includegraphics[width=0.82\linewidth]{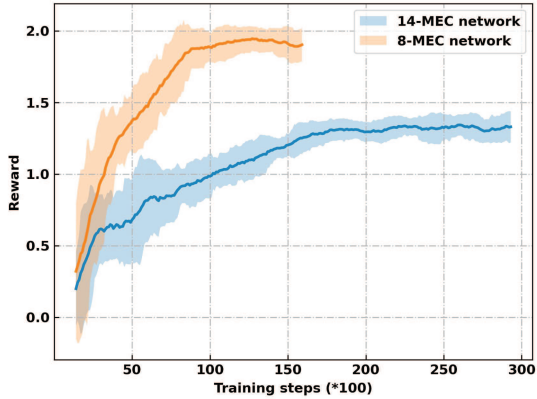}
    \caption{The convergence property of DRL algorithms in two networks.}
    \label{convergence}
\end{figure}

Figure \ref{time-cost} displays the running time of our proposed solutions for different traffic requirements using a 3.2GHz 6‑core 8th‑generation Intel Core i7 machine. In the case of full traffic, MILP takes 544s to obtain the optimal result, thereby violating the real-time response requirements in 5G services. In contrast, the DRL-based strategy exhibits a reasoning time of less than 1 second, highlighting its significance in practical applications.

\subsection{14-MEC network}
To further explore the performance of the DRL-based solution in a larger network, we assume Node 2 and Node 9 are pre-activated and simulate the locations of all provisioned functions and routing paths for another traffic set $T$ with 14 requests in network \ref{large}. Traffic set $T$ consists of fronthaul delay [14, 20, 17, 19, 19, 17, 17, 17, 13, 15, 20, 18, 16, 11] united in km, end-to-end delay [28, 38, 40, 29, 41, 32, 34, 42, 37, 29, 28, 31, 28, 31] united in km, data size [11, 9, 8, 9, 9, 8, 10, 11, 10, 9, 8, 10, 9, 9] united in Gbits/s and computing resource requirements for DU and UPF [11, 9, 8, 9, 9, 8, 10, 11, 10, 9, 8, 10, 9, 9] and [3, 4, 5, 5, 3, 3, 5, 5, 3, 3, 3, 3, 5, 3] united in cores. These results are illustrated in Figure \ref{drl-14}, where standby and newly activated MECs are marked by blue and green circles, respectively. The red arrows indicate the selected routing paths. 

Similar to the 8-MEC network, we also compare the energy consumption of $T$ under various management policies as network idle time increases, as demonstrated in Figure \ref{vacant-14}. Due to the maintenance costs of pre-activated nodes, the energy costs of all strategies rise as idle time increases. In contrast, the energy costs of GHP, PMD and DRL would remain constant, as observed in Figure \ref{vacant-8}.

Figure \ref{150-14} showcases the energy cost of 50 groups of random requests, each with different parameter configurations, observed over a 150s network idle period. The DRL-based solution maintains its advantage over other strategies and effectively addresses energy consumption challenges in larger networks. However, it is important to note that as the network size grows, the relative benefits obtained by DRL gradually diminish. This trend might be attributed to the escalating complexity of inter-agent cooperation with the growth in agent count, which in turn potentially compromises the precision of the algorithm's fitting performance. To address the challenges of larger network scales, a possible solution is to segment these networks into smaller sub-networks or clusters based on the geographical distribution of MECs. By independently applying the DRL algorithm to each segmented cluster, this solution can both simplify network management and compensate for the limitation of DRL in managing expansive networks.

\subsection{DRL training cost}
In this subsection, we analyze the time complexity, space complexity, and execution cost of MADDPG in different network scenarios. Referring to \cite{leem2020action}, the time complexity of reinforcement learning is sub-linear in the length of the state period and can be represented as $O$(training steps). To illustrate this, we present the convergence properties of MADDPG in Figure \ref{convergence}. The yellow and blue lines represent the convergence performance in the 8-MEC network with 2 agents and the 14-MEC network with 3 agents, respectively. The faster convergence speed and improved reward of DRL in the smaller network can be partially attributed to its fewer agents and reduced network complexity in terms of state and action spaces.

In the smaller network, two agents share the responsibility of activation management, with each handling 4 nodes. On the other hand, in the larger network, three agents divide the responsibility, with each managing 5, 5, and 4 nodes, respectively. Furthermore, the space complexity has been proven to be sub-linear in the size of state space, action space, and step numbers per episode. It can be expressed as $O(\sum_l^L\sum_q^Q F_{s}^{lq}F_{a}^{lq}F_{h}^{lq})$, where $Q$ is the set of actors and critics, $F_s$ denotes the number of states, $F_a$ indicates the number of outputs, and $F_h$ represents the number of training steps in each episode. Based on the values above, the space complexities of MADDPG in the small and large networks can be calculated as $O(2.84E5)$ and $O(1.21E6)$, respectively. Therefore, DRL algorithm may incur significant energy consumption during its training phase, potentially offsetting its energy-saving advantages \cite{strubell2019energy, garcia2019estimation}. However, this can be viewed as a one-time overhead. Once the model is trained, it can continuously serve the network, offering sustainable energy savings.
In summary, with acceptable complexity, the DRL-based solution approaches the effectiveness of MILP and contributes to energy savings in baseband function deployments for Open RAN.

\section{Feasibility Evaluation over Open RAN Testbed}
\label{testbed}

Leveraging OpenDaylight, OpenStack, OSM, building upon the frameworks proposed in \cite{d2022orchestran, polese2023understanding}, an innovative Open RAN testbed is designed and implemented to validate the energy-saving performance of our proposed algorithms.
The following subsections provide a comprehensive description of the testbed orchestration and implementation, grounded in the Open RAN standards. Furthermore, we explored the activation costs of a MEC server at three distinct depth hibernation levels, its loading power footprint, and the transmission delay between MECs.

\begin{figure}[b]
    \centering
    \includegraphics[width=0.87\linewidth]{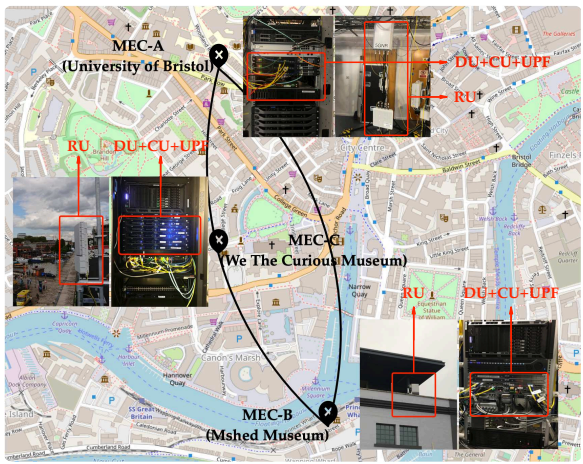}
    \caption{Open RAN testbed urban deployment in Bristol}
    \label{testbed0}
\end{figure}

\begin{figure}[t]
    \centering
    \includegraphics[width=0.87\linewidth]{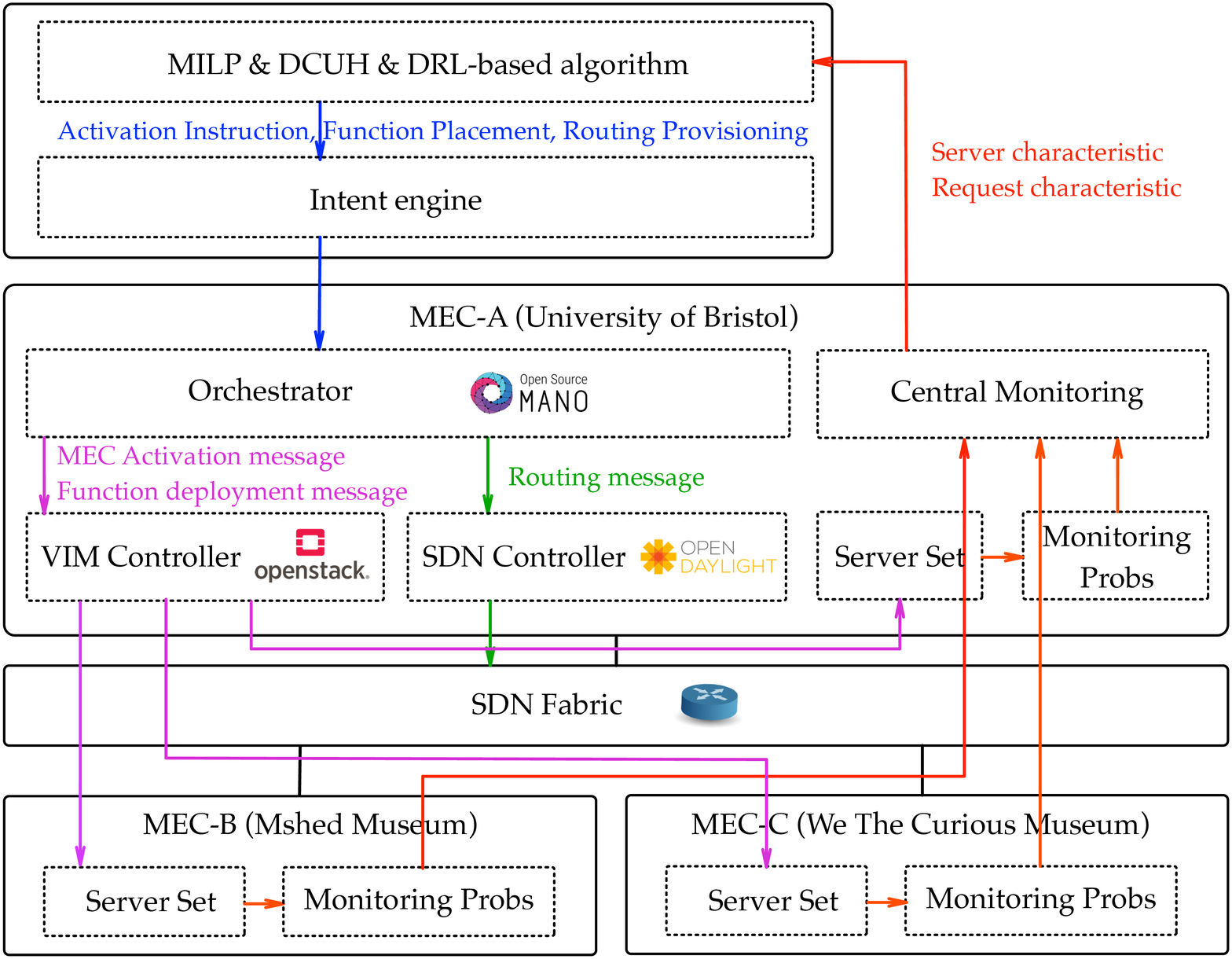}
    \caption{Closed-loop Open RAN Testbed Orchestration.}
    \label{testbed1}
\end{figure}

\subsection{Testbed Orchestration}
The Open RAN testbed deployment and its orchestration are shown in Figures \ref{testbed0} and \ref{testbed1}. There are three MECs distributed at the University of Bristol (MEC-A, main), MShed Museum (MEC-B) and We The Curious Museum (MEC-C). Each of them consists of a server set and several edge monitoring probes. The RAN Intelligent Controller (RIC) and Service Management and Orchestration (SMO) are integrated into the testbed to facilitate the administration and optimization of network resources and services. 
RIC (realized based on Juniper RIC \cite{Juniper2023}) plays a vital role in enhancing the performance and functionality of the RAN. It hosts our proposed Python-based algorithm in the container as xApps and collaborates with the MEC platform to govern server activation and function placement decisions.
Within RIC, an Intent Engine (IE) that converts Python script output to OSM standard YAML file \cite{mcnamara2023nlp} is incorporated to establish closed-loop automation, encompassing intent capture, intent translation, and virtualized function activations, which consistently monitor and adjust to assure service alignment with end-to-end requirements.
In addition, SMO is integrated with the Virtualized Infrastructure Management (VIM) controller and the Software-Defined Network (SDN) controller to execute path provisioning and baseband function deployment, respectively. These control components are placed on MEC-A. To get the DRL state including server activation status and resource information, we build a REST API to expose the information collected by Monitoring Prods from the server set to the Central Monitoring on MEC-A.

\begin{figure}[t]
    \centering
    \includegraphics[width=0.85\linewidth]{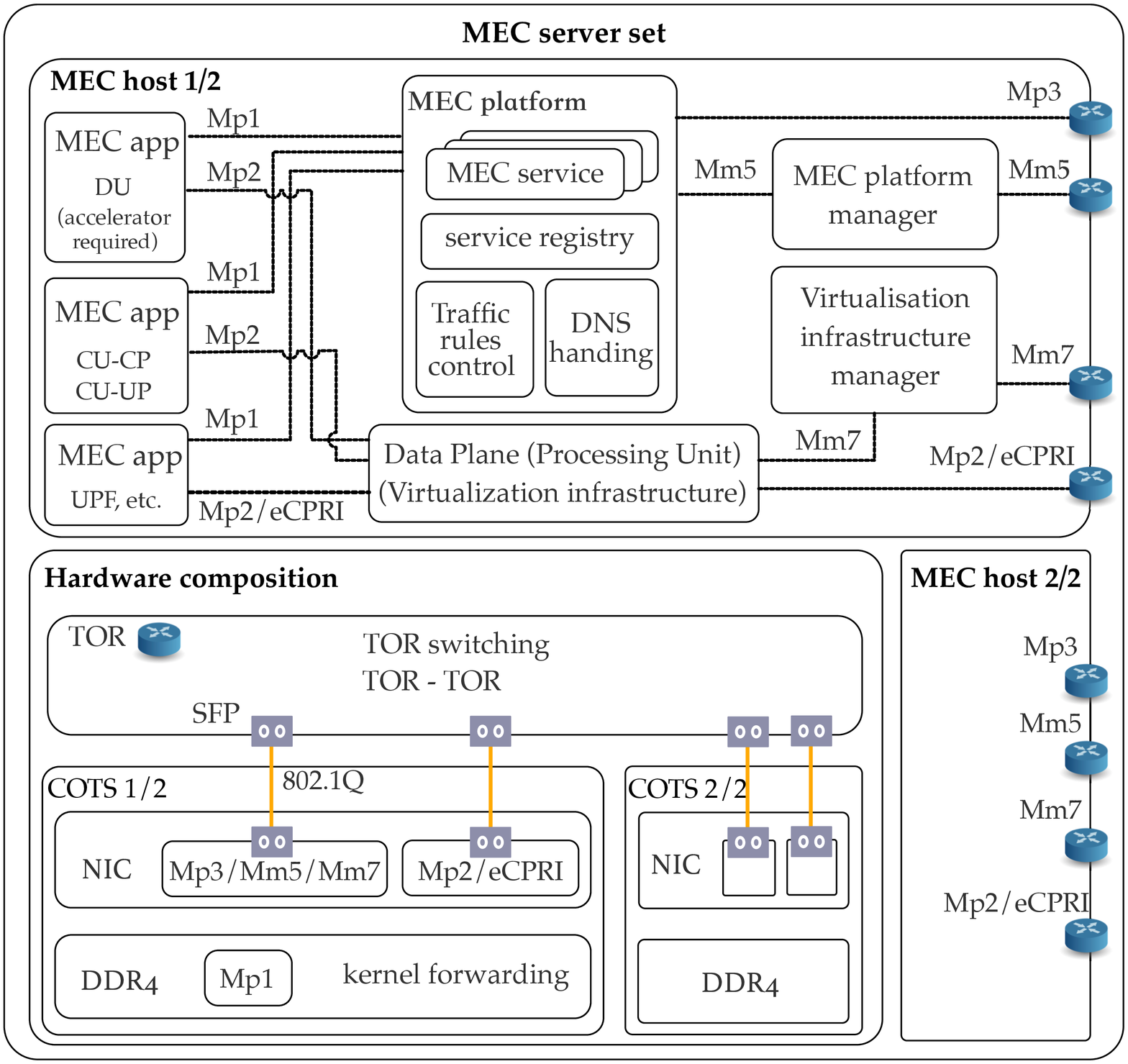}
    \caption{Baseband function deployment based on ETSI and Open RAN standards in a MEC server set. (Yellow lines denote wired transmission based on 802.1Q; Gray squares denote the SFP transceiver; Blue cylinders indicate network switching).}
    \label{testbed2}
\end{figure}

In addition, the MEC architecture depicted in Figure \ref{testbed2} demonstrates the implementation of baseband functions in our testbed in accordance with the European Telecommunications Standards Institute (ETSI) and Open RAN standards. The MEC server set comprises a top-of-rack (TOR) switch and two commercial-off-the-shelf (COTS) components, providing in-host switching and external connectivity. In particular, in-host switching is facilitated by Linux kernel packet switching and forwarding packages, while external connectivity is achieved through the network interface controller (NIC), which connects to the TOR switch via fiber using small form-factor pluggable transceivers (SFPs). Consequently, the TOR switch can perform both TOR and TOR-TOR switching within the rack and between MECs. It is worth mentioning that all connections to the TOR switch, indicated by yellow lines, are implemented using fiber and IEEE 802.1Q protocols \cite{IEEE}.

These physical devices are the basis for virtualized MEC host modules and interfaces. DU, CU control plane (CU-CP), CU user plane (CU-UP), and UPF are implemented as MEC applications (MEC apps). In specific, DU and CU functions are achieved by the open-source srsRAN platform \cite{johnson2022nexran}, and UPF is executed by Open 5GS \cite{neto2021analysis}.
Furthermore, within a single COTS component, two NICs are designated for O2, E2 and A1 interfaces, wherein O2 enables the communication between different MECs in our testbed, therefore, supports coordination and collaboration between MEC instances. 
E2 manages the transmission between baseband functions.
A1 interface allows the RAN Intelligent Controller (RIC) to manage the hibernation, activation, or standby states of the MEC apps. In addition, the kernel forwarding function utilizing non-uniform memory access (NUMA) on the installed DDR4 memory modules can be considered as O1 interface, which links the MEC platform with MEC apps. 
The switches connected by A1, O2 and E2 interfaces on the right side of the MEC host demonstrate that all external communications between MEC hosts must pass through the TOR switch.

\subsection{Feasibility verification}
\begin{figure}[t] 
    \centering  
        \setlength{\belowcaptionskip}{-0.2cm} 
    \subfigure[Power variation of startup]{
        \label{All_1}
        \includegraphics[width=0.482\linewidth]{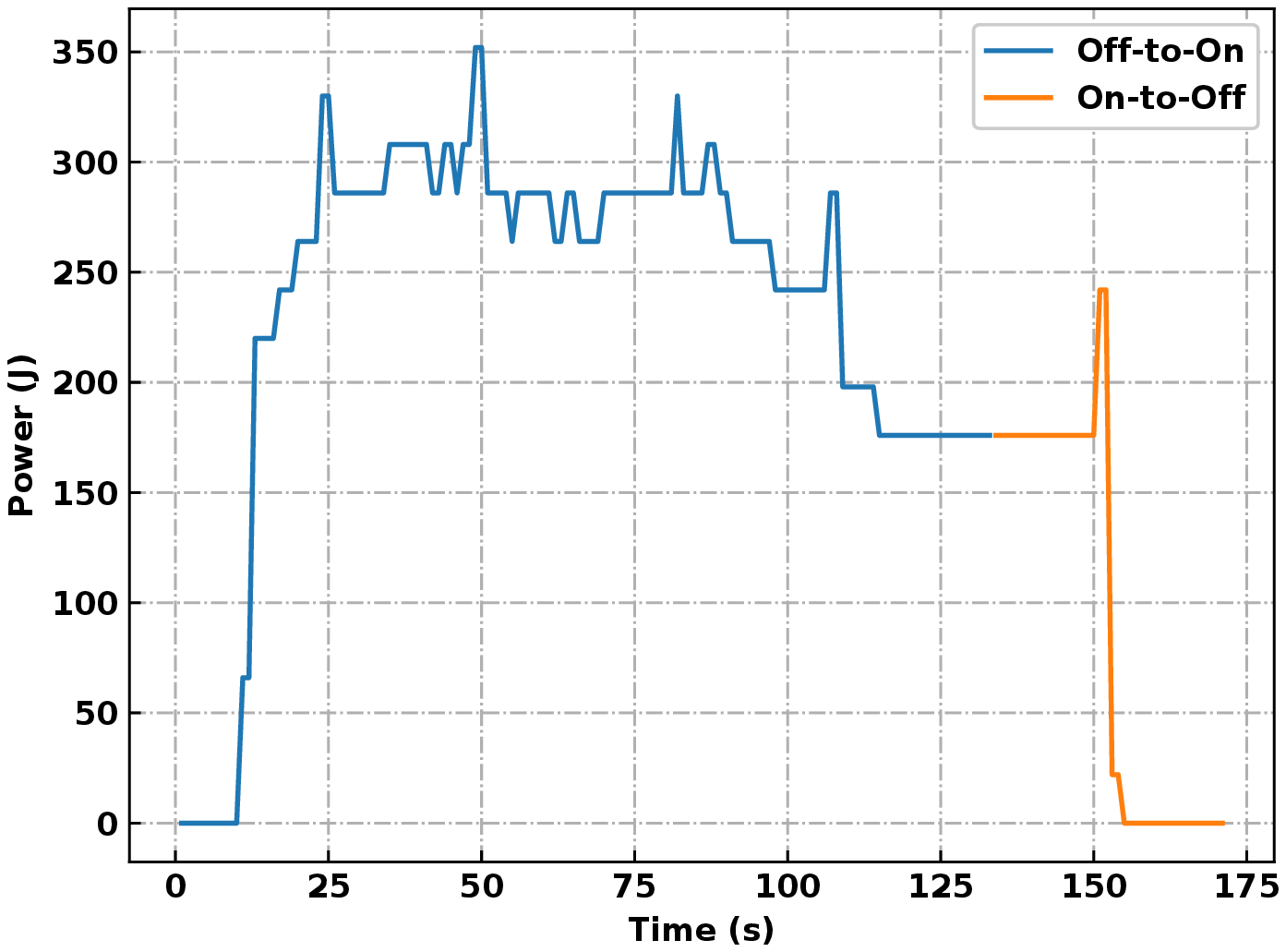}}
    \subfigure[Temperature Variation of Startup]{
        \label{All_2}
        \includegraphics[width=0.477\linewidth]{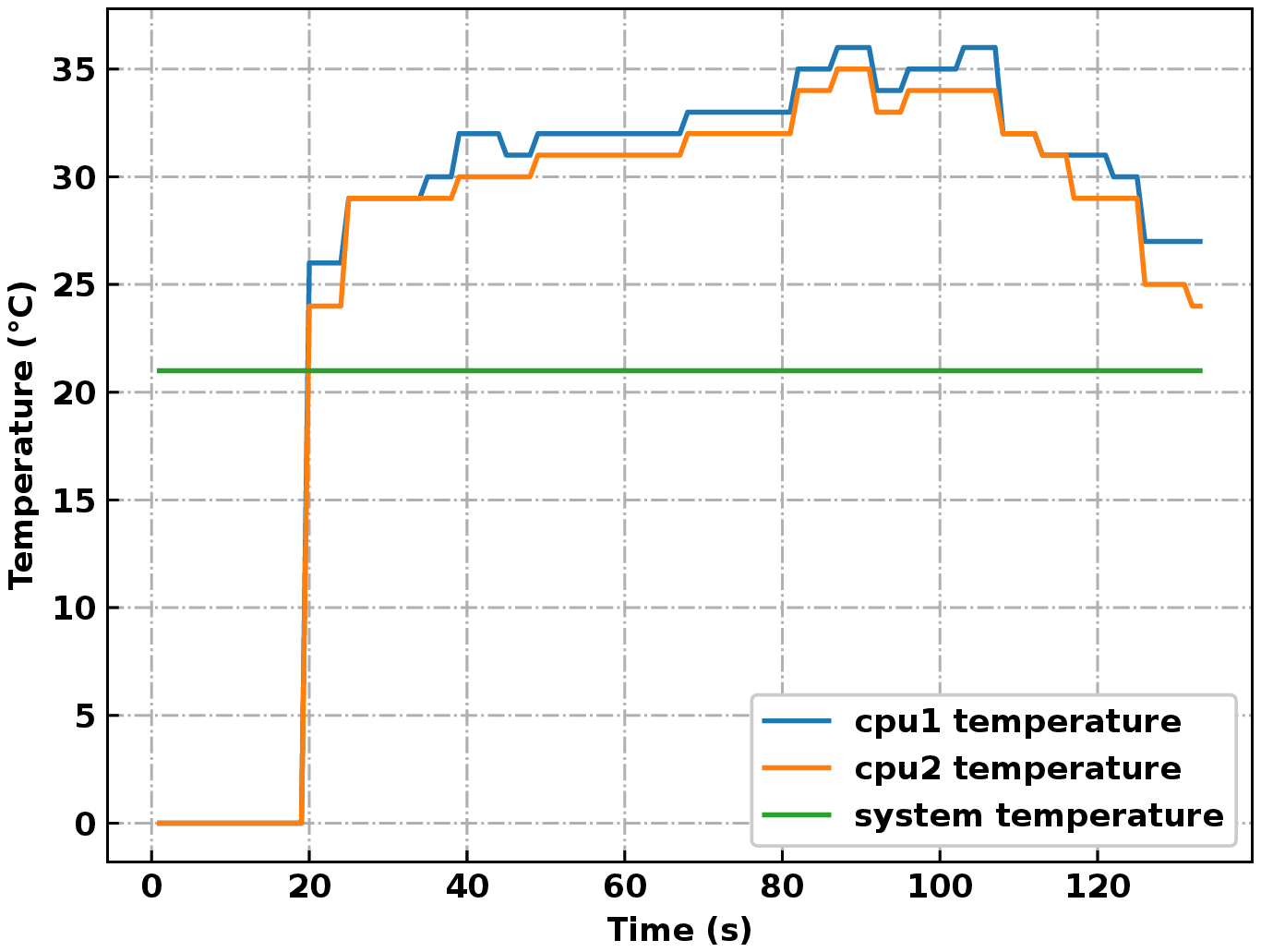}}
    \subfigure[Power Variation of Cold Reboot]{
        \label{All_3}
        \includegraphics[width=0.472\linewidth]{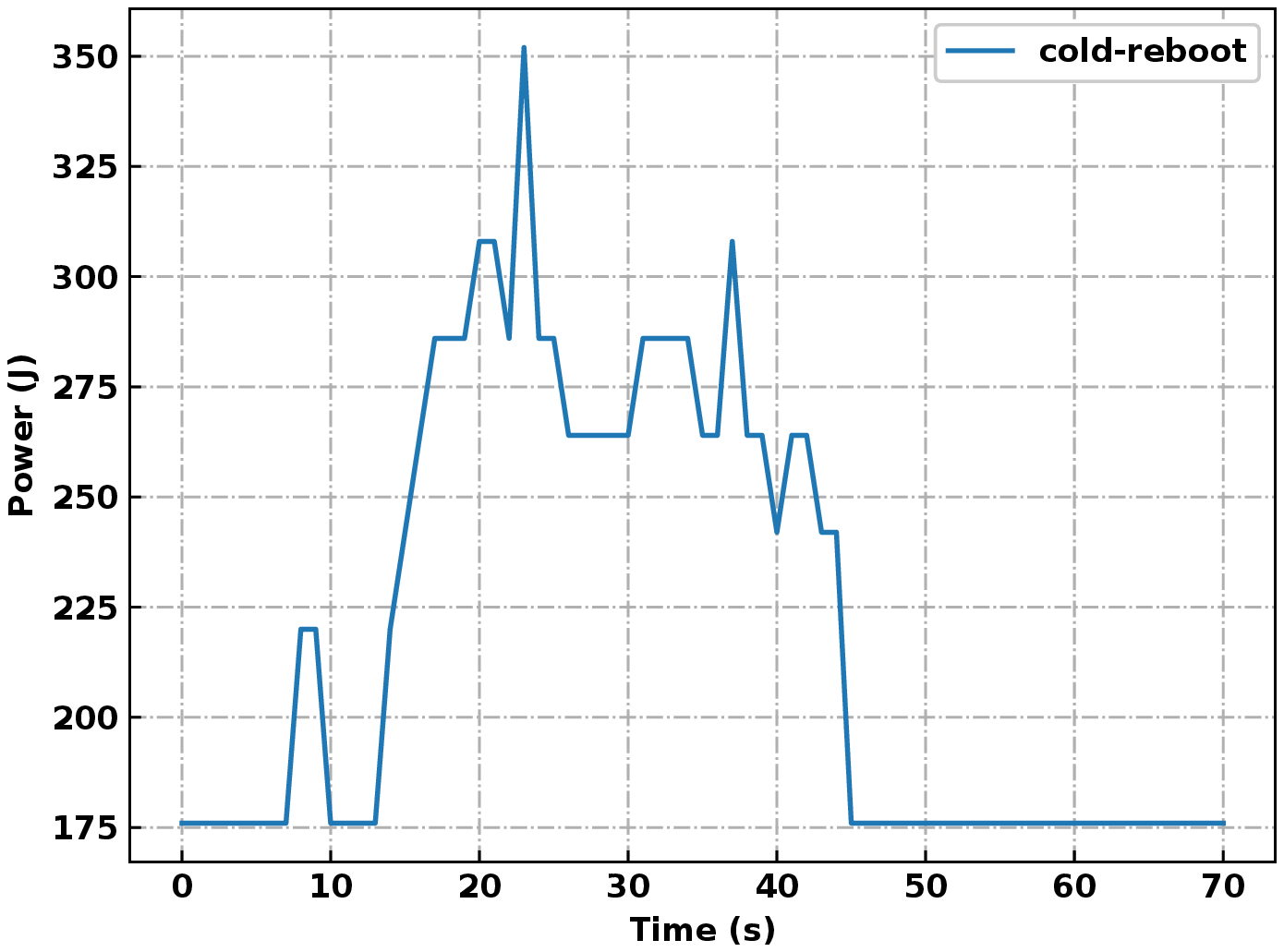}}
    \subfigure[Temperature Variation of Cold Reboot]{
        \label{All_4}
        \includegraphics[width=0.478\linewidth]{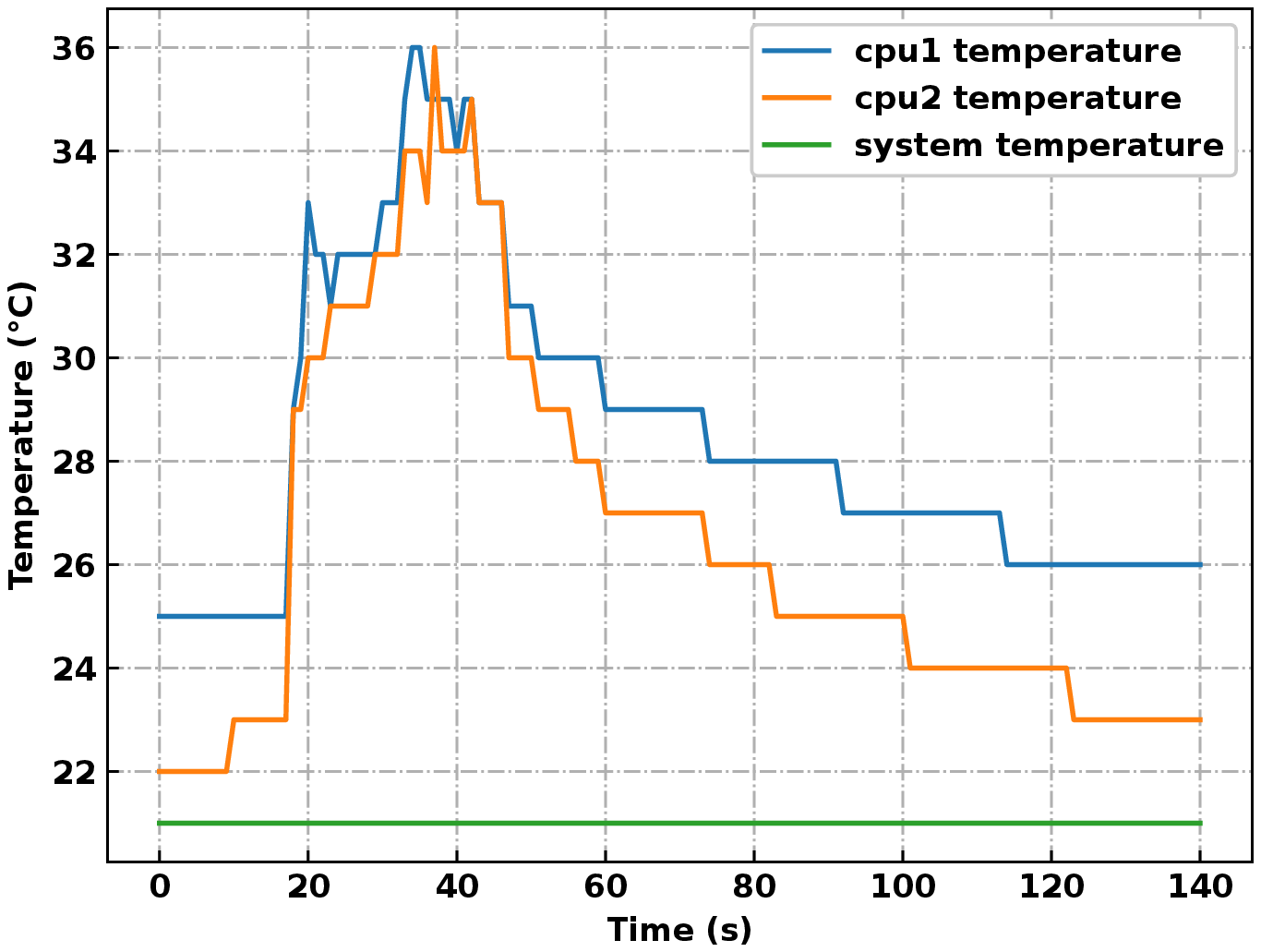}}
    \subfigure[Power Variation of Warm Reboot]{
        \label{All_5}
        \includegraphics[width=0.475\linewidth]{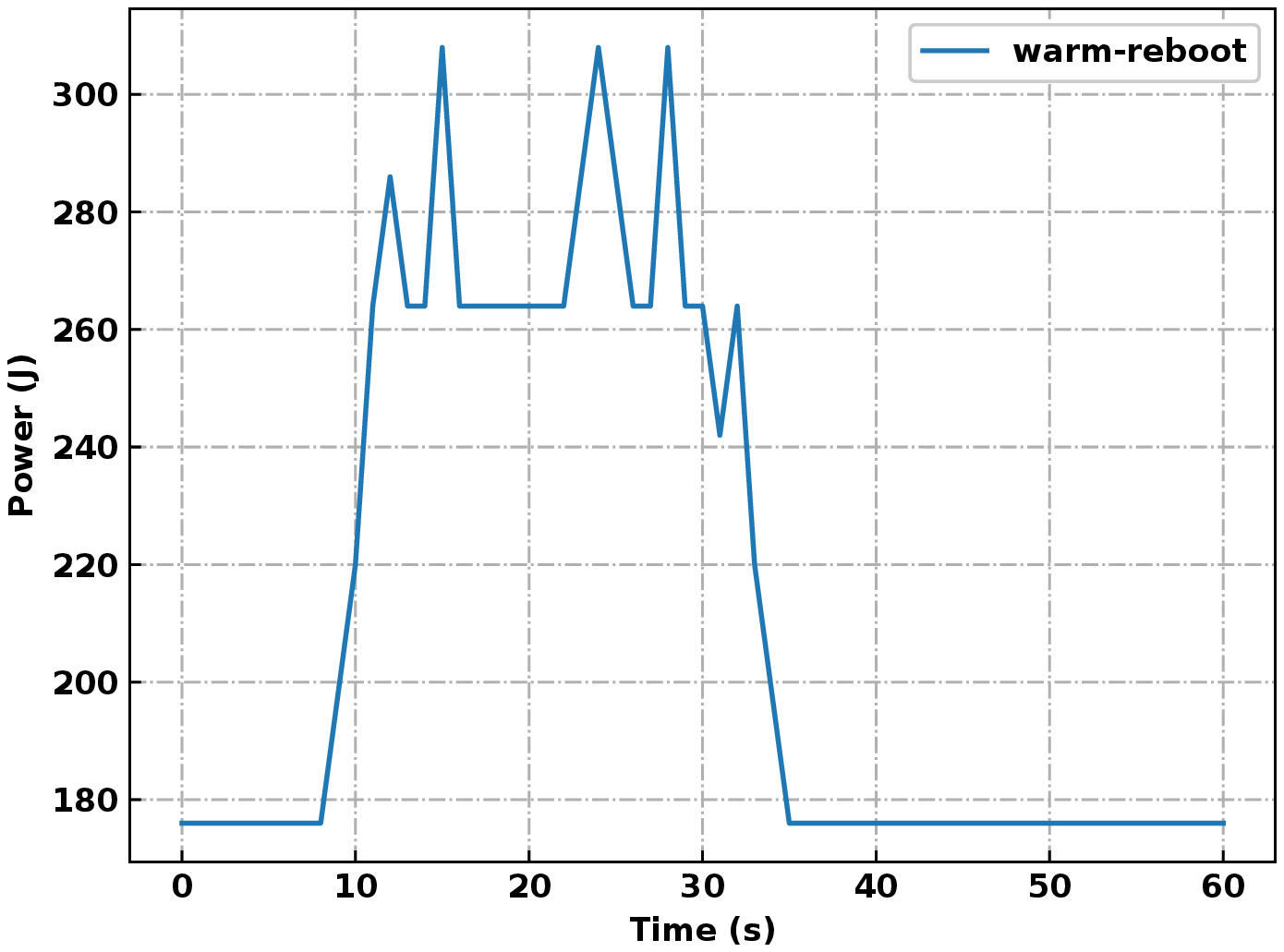}}
    \subfigure[Temperature Variation of Warm Reboot]{
        \label{All_6}
        \includegraphics[width=0.48\linewidth]{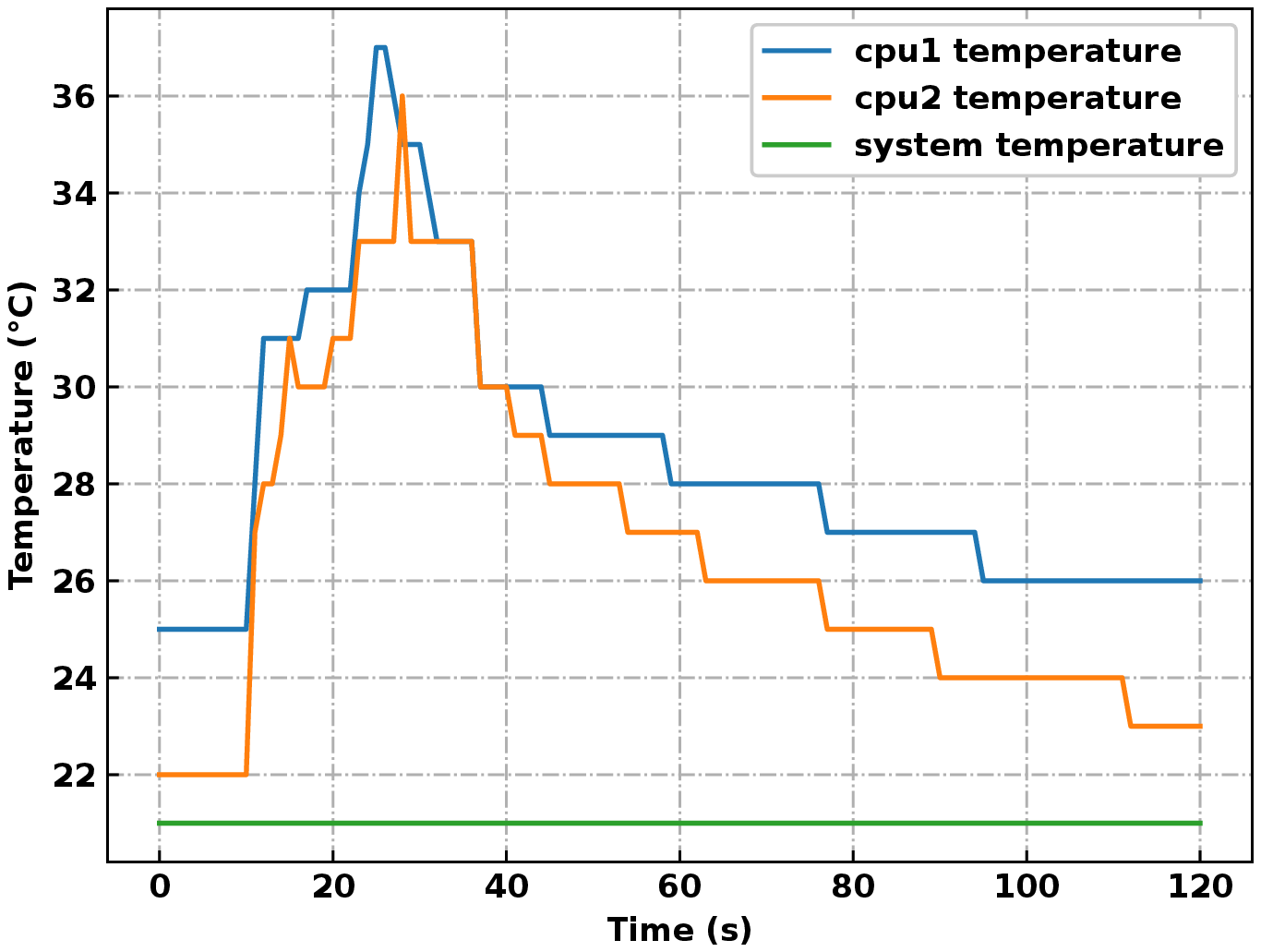}}
    \caption{Power and temperature variation of a server during three types of server activation.}
	\label{All}
\end{figure}

To demonstrate the importance of activation cost in determining the effectiveness of baseband function deployment algorithms, the activation time and corresponding power, CPU temperature, and system temperature changes of MEC-A during startup, cold reboot, and warm reboot processes are explored in this subsection. The examined server has 2 CPUs, each with 14 cores and 2 threads per core. A startup wakes up the server from the shutdown state. A cold boot resets running hardware and reloads the operating system. A warm boot, on the other hand, regains the initial state of a server without hampering the power source. 
As shown in Figure \ref{All}, MEC experiences a manifest power surge for a certain period before returning to a stable state.
Moreover,  it is observed that the more profound the dormancy degree of the server, the longer and higher the activation time and the activation power it needs. 
Activation energies for startup, cold, and warm reboots are approximately 26.9kJ, 8.1kJ, and 7.1kJ, respectively. Therefore, maintaining idle servers can save more energy during heavy network traffic. In addition, these results reaffirm our assertion that energy, instead of power, should be the metric of choice when evaluating baseband function management policy, as it considers the span of activation time. 
Additionally, we examined the inference time on the testbed, which encompasses the state information collection delay (from Monitoring Probs to Central Monitoring), the DRL solving time (xApp operation duration), and the policy execution delay (from RIC on MEC-A to MEC-B and MEC-C). The total inference time is less than 200ms on our testbed.

In summary, the orchestration of the Open RAN testbed and test result prove the feasibility of management algorithms in real networks and serve as a base for future studies in designing more practical baseband function deployment strategies.
\vspace{-0.1cm}
\section{Conclusion and future work}
This paper addresses the urgent challenges associated with effectively managing baseband function deployment within the burgeoning Open RAN architecture, where a greater distribution of MECs and time-space usage dynamics pose significant burdens on network energy-saving efficiency. 
Conventional strategies exhibit shortcomings due to the insufficient consideration of activation energy costs while managing server activation status.
In this paper, under the framework of Open RAN, a robust multi-agent DRL-based algorithm is devised to minimize energy consumption over the network, subject to resource and latency constraints of servers and requests.
A state-of-the-art Open RAN testbed is also prototyped to demonstrate the validity and feasibility of the proposed algorithm.
Simulation results illustrate the superior performance of our DRL-based solution, which approaches the benchmark of MILP and significantly outperforms GHP and other existing strategies. By emphasizing the importance of considering multiple UPFs and the activation time of servers, this work represents a significant advancement in enhancing energy-saving efficiency within Open RAN systems. 

Regarding future work, there is room to refine the energy model by accounting for the nonlinear associations between request size and server power consumption. Furthermore, enhancing the performance of MADDPG in the face of 5G network structural dynamics presents another avenue for improvement. In light of these challenges, we intend to conduct a loading stress test on our testbed servers and use the findings to establish a more precise system model. Concurrently, we aim to explore the creation of generalized DRL models to accommodate diverse network structures.

\section*{Acknowledgements}
The authors would like to express their gratitude for the support from the China Scholarship Council and the UK-funded project REASON under the Future Open Networks Research Challenge sponsored DSIT.

\bibliographystyle{IEEEtran} \bibliography{references}

\end{document}